\documentclass{article}
\usepackage{url}
\usepackage[utf8]{inputenc}
\usepackage{fullpage}
\usepackage{amsfonts,amssymb,amsmath}
\usepackage{xcolor}
\usepackage{dsfont}
\usepackage{caption}
\usepackage{color}
\usepackage{graphicx}

\newcommand{\Ll}{L}
\newcommand{\A}{A}
\newcommand{\D}{D}
\newcommand{\B}{B}
\newcommand{\F}{F}
\newcommand{\C}{C}

\newcommand{\x}{\mathbf{x}}
\newcommand{\y}{\mathbf{y}}
\newcommand{\f}{\mathbf{f}}
\newcommand{\e}{\mathbf{e}}
\newcommand{\h}{\mathbf{h}}

\newtheorem{definition}{Definition}
\newtheorem{theorem}{Theorem}
\newtheorem{proposition}{Proposition}

\newtheorem{assumption}{Assumption}
\newtheorem{corollary}{Corollary}

\usepackage[round]{natbib}
\DeclareMathOperator{\var}{var}

\title{Granger Causality: A Review and Recent Advances}
\author{Ali Shojaie and Emily B. Fox \\
University of Washington} 
\date{}

\begin{document}
\maketitle

\begin{abstract}
Introduced more than a half century ago, Granger causality has become a popular tool for analyzing time series data in many application domains, from economics and finance to genomics and neuroscience. Despite this popularity, the validity of this notion for inferring causal relationships among time series has remained the topic of continuous debate. Moreover, while the original definition was general, limitations in computational tools have primarily limited the applications of Granger causality to simple bivariate vector auto-regressive processes or pairwise relationships among a set of variables. Starting with a review of early developments and debates, this paper discusses recent advances that address various shortcomings of the earlier approaches, from models for high-dimensional time series to more recent developments that account for nonlinear and non-Gaussian observations and allow for sub-sampled and mixed frequency time series. 
\end{abstract}

\section{INTRODUCTION}\label{sec:intro}
There is a range of applications where the interest is in understanding interactions between a set of time series, including in neuroscience, genomics, econometrics, climate science, and social media analysis.  For example, in neuroscience, one may seek to understand whether activity in one brain region correlates with later activity in another region, or to decipher \emph{instantaneous} correlations between regions---both notions of \emph{functional connectivity}.  In genomics, there is an analogous study of gene regulatory networks.  In econometrics, one may be interested in how various macroeconomic indicators predict one another. We also have unprecedented levels of data on people's actions---whether they be social media posts, purchase histories, or political voting records---and want to understand the dependencies between the actions of these individuals. Modern recording modalities and the ability to store and process large amounts of data have escalated the scale at which we seek to do such analyses.  

In many cases, one may seek notions of \emph{causal} interactions amongst the time series, but be limited to drawing inferences from observational data without opportunities for experimentation and without known mechanistic models for the observed phenomena.  In such cases, \citet{granger1969} put forth a framework leveraging the temporal ordering inherent to time series in hopes of drawing causal statements restricted to the past causing the future.  The framework, in reality, assesses whether one series is \emph{predictive} of another: a series $x_i$ is deemed \emph{not} to be `causal' of another series $x_j$ if leveraging the history of series $x_i$ does not reduce the variance of the prediction of series $x_j$.  In this review, we distinguish this definition from other standard definitions of causality by referring to it as \emph{Granger causality}.  Although there is a long history of debate about the validity of the Granger causality framework for causal analyses---and justly so---in this review we take the stance that analyzing interactions in time series defined by association has its utility.  

Granger causality has traditionally relied on assuming a linear vector autoregressive (VAR) model~\citep{lutkepohl2005} and considering tests on the VAR coefficients in the bivariate setting. However, in real-world systems involving many time series, considering the relationship between just a pair of series can lead to confounded inferences\citep[ e.g.,][]{lutkepohl1982}.  \emph{Network} Granger causality aims to adjust for possible confounders or jointly consider multiple series~\citep{eichler2007,basu2015ngc}.  There are other important limitations of the linear VAR model underlying standard Granger causal analysis that have precluded its broad utility.  Some limiting assumptions include assuming 
(i) real-valued time series with (ii) linear dynamics dependent on a (iii) known number of past lagged observations, with (iv) observations available at a fixed, discrete sampling rate that matches the time scale of the causal structure of interest.  In contrast, modern time series are often messy in ways that break a number of these assumptions, including through non-linear dynamics and irregular sampling.  Recent advances have pushed the envelope on where Granger causality can be applied by loosening these restrictions in a variety of ways.  We review some of these advances, and set the stage for further developments.  

\subsection{Outline of Review}
In Section~\ref{sec:classics} we review the history of Granger causality, starting with the original definition and assumptions in Section~\ref{sec:gcdefinition} and early approaches for testing in Section~\ref{sec:earlymethods}. We then turn to network Granger causality and the issues of lag selection and non-stationary VAR models in Section~\ref{sec:ngc}.  Finally, in Section~\ref{sec:general} we review recent advances that move beyond the standard linear VAR model and consider discrete-valued series (Section~\ref{sec:categorical}), non-linear dynamics and interactions (Section~\ref{sec:nonlinear}), and series observed at different sampling rates (Section~\ref{sec:mixedFreqGranger}).

\section{THE HISTORY OF GRANGER CAUSALITY} 
\label{sec:classics}

\subsection{Definition}\label{sec:gcdefinition}
In his seminal paper, \citet{granger1969} proposed a notion of `causality' based on how well past values of a time series $y_t$ could \emph{predict} future values of another series $x_t$. 
Let $\mathcal{H}_{<t}$ be the history of \emph{all relevant information} up to time $t-1$, and $\mathcal{P}(x_t \mid \mathcal{H}_{<t})$ the \emph{optimal prediction} of $x_t$ given $\mathcal{H}_{<t}$. 
Granger defined $y$ to be `causal' for $x$ if 
\begin{equation}\label{eqn:gcdef}
    \var\big[x_t - \mathcal{P}(x_t \mid \mathcal{H}_{<t})\big] <     \var\big[x_t - \mathcal{P}(x_t \mid \mathcal{H}_{<t}\backslash y_{<t})\big],
\end{equation}
where $\mathcal{H}_{<t}\backslash y_{<t}$ indicates excluding the values of $y_{<t}$ from $\mathcal{H}_{<t}$.
That is, the variance of the optimal prediction error of $x$ is reduced by including the history of $y$ (informally, $y$ is `causal' of $x$ \emph{if past values of $y$ improve the prediction of $x$}). 
This characterization is clearly based on \emph{predictability} and does not (directly) point to a causal effect of $y$ on $x$: $y$ improving the prediction of $x$ does not mean $y$ \emph{causes} $x$. 
Nonetheless, assuming causal effects are ordered in time (i.e., cause before effect), Granger argued that, \emph{under some assumptions}, if $y$ can predict $x$, 
then there must be a mechanistic (i.e. causal) effect; that is, predictability implies causality. 
We explicitly refer to this definition as \emph{Granger causality} throughout this review to distinguish it from other formal definitions of causality.

While the definition seems general and does not rely on specific modeling assumptions, Granger's original argument was based on the identifiability of a unique \emph{linear} model. 
Denoting the vector of variables at time $t$ by $\x_t = (x_{1t}, x_{2t}, \ldots, x_{pt})^\top$, he considered the linear model 
\begin{equation}\label{eqn:linmodel}
    A^0 \x_t = \sum_{k=1}^d A^k \x_{t-k} + \e_t,  
\end{equation}
where $A^0, A^1, \ldots, A^d$ are $p \times p$ \emph{lag matrices} (coefficients) and $d$, the \emph{lag} or \emph{order}, may be finite or infinite. 
The $p$-dimensional white noise innovation, or error, term $\e_t$ can have a diagonal or non-diagonal covariance matrix $\Sigma$. 

\citet{granger1969} pointed out that this model is generally \emph{not identifiable} (the matrices $A^k$ are not uniquely defined) unless $A^0$ is diagonal. 
Granger referred to this special case---corresponding to the well-known \emph{vector auto-regressive} (VAR) model \citep{lutkepohl2005}---as a ``simple causal model'', distinguishing it from models with \emph{instantaneous causal effects} when $A^0$ has nonzero off-diagonal entries. This more general form of (\ref{eqn:linmodel}) is known as a \emph{structural vector auto-regressive} (SVAR) model \citep{kilian2013svar}, and can be identified under certain parameter restrictions \citep{kilian2017structural}. Such SVAR are further considered in Section~\ref{sec:mixedFreqGranger}.

The model in (\ref{eqn:linmodel}) is clearly restrictive and does not (dis)prove the presence of causal effects. In particular, there are a number of implicit and explicit restrictive assumptions required for the (S)VAR model to be an appropriate framework for identifying Granger causal relationships:
\begin{itemize}
    \item \textbf{Continuous-valued series} All series are assumed to have continuous-valued observations. However, many interesting data sources---such as social media posts or health states of an individual---are discrete-valued. 
    \item \textbf{Linearity} The true data generating process, and correspondingly the causal effects of variables on each other, is assumed to be linear. In reality, many real-world processes are non-linear. 
    \item \textbf{Discrete-time} The sampling frequency is assumed to be on a discrete, regular grid matching the true causal time lag. If the data acquisition rate is slower or otherwise irregular, causal effects may not be identifiable. Likewise, the analysis of point processes or other continuous-time processes is precluded.
    \item \textbf{Known lag} The (linear) dependency on a history of lagged observations is assumed to have a known order.  Classically, the order was not estimated and taken to be uniform across all series.
    \item \textbf{Stationarity} The statistics of the process are assumed time-invariant, whereas many complex processes have \emph{evolving} relationships (e.g., brain networks vary by stimuli and user activity varies over time and context).
    \item \textbf{Perfectly observed} The variables need to be observed without \emph{measurement errors}. 
    \item \textbf{Complete system} All relevant variables are assumed to be observed and included in the analysis, i.e., there are \emph{no unmeasured confounders}. This is a stringent requirement, especially given that early approaches for Granger causality focused on the bivariate case---that is, they did not account for \emph{any} potential confounders. 
\end{itemize}

The above requirements were discussed in Granger's original and followup papers \citep{granger1969, granger1980, granger2001} and extensively by other authors \citep{stokes2017study, maziarz2015review}; see also the recent review by \citet{glymour2019}. 
Unfortunately, each of the above requirements is unlikely to hold in practice. These assumptions are also not verifiable, and even more unlikely to hold simultaneously, which is what is required for the identifiability of causal effects. In fact, Granger admitted this limitation and gave examples of cases where causal effects could not be identified or wrong conclusions could be drawn. However, in each case, he presented an argument for why the example did not violate the basic principle, either by giving justifications through an alternative model \citep{granger1988some}, or by adding disclaimers (e.g. the definition cannot be applied to deterministic or perfectly predictable processes). 

The debate over the notion of `causality' introduced by Granger has continued since its introduction. An illustrative example is the commentary by \citet{sheehan1982} who used Granger causality to show that the U.S. Gross National Product (GNP) \emph{causes} sunspots; the rebuttal by \citet{noble1983} suggests an alternative model would have led to a different conclusion.
Despite its limitations, \citet{granger1980} and a number of other researchers, including prominent econometricians \citep{sims1972, bernanke1992}, have argued that the approach can be used to identify causal effects. Researchers in various applied domains, from neuroscience \citep{bergmann2021inferring, reid2019advancing} to environmental sciences \citep{cox2015has}, have used Granger's framework to (informally) draw causal conclusions. 
Other researchers have emphasized the limitations of the approach and have tried to distinguish it as \emph{Granger causality} or \emph{G-causality} \citep{holland1986, bressler2011wiener}. 

While limited and not generally informative about causal effects, the notion of Granger causality can lead to useful insights about interactions among random variables observed over time. 
In the next section, we discuss early approaches for identifying Granger causality and its applications in various domains. In the remaining sections, we discuss approaches that aim to (partially) address some of the limitations of the original Granger causality framework and relax some of the requirements discussed above.

\subsection{Early approaches and applications}
\label{sec:earlymethods}

The basic definition (\ref{eqn:gcdef}) requires that \emph{all} relevant information is accounted for when testing whether series $x$ Granger causes series $y$. However, early methods for identifying Granger causality were limited to bivariate models, ignoring the effect of other variables.
In his original paper, \citet{granger1969} used an argument based on spectral representation, using \emph{coherence} and \emph{phase}, to motivate the original definition. 
Using a bivariate version of the SVAR model (\ref{eqn:linmodel}) (i.e., with $p=2$), he then showed that when $A^0$ is diagonal (i.e., a simple causal model / VAR model), Granger causality corresponds to nonzero entries in the autoregressive coefficients.  In particular, for a bivariate model
\begin{equation}
    \label{eqn:bivariate}
\begin{aligned}
    a_x^0 x_t &= \sum_{k=1}^d a^k_{xx} x_{t-k} + \sum_{k=1}^d a^k_{xy} y_{t-k} + e_{t,x}\\
    a_y^0 y_t &= \sum_{k=1}^d a^k_{yy} y_{t-k} + \sum_{k=1}^d a^k_{yx} x_{t-k} + e_{t,y},
\end{aligned}
\end{equation}
\citet{granger1969} concluded that \emph{series $y$ is Granger causal for series $x$ if and only if $a^k_{xy} \ne 0$ for some $1 \le k \le d$}.

\citet{sims1972} later gave an alternative definition of Granger causality based on coefficients in a moving average (MA) representation. 
The characterizations by \citet{granger1969} and \citet{sims1972}, which have been shown to be equivalent \citep{chamberlain1982}, can be tested using an $F$-test comparing two models: the \emph{full model}, including past values of both $x$ and $y$, and the \emph{reduced model}, including only past values of $x$. Formally, 
\begin{equation}\label{eqn:Ftest}
F = 
\frac{\left(\text{RSS}_{\text{red}}-\text{RSS}_{\text{full}}\right)/(r - s)}{\text{RSS}_{\text{full}}/(T-r)},
\end{equation}
where $\text{RSS}_{\text{full}}$ and $\text{RSS}_{\text{red}}$ are the residual sum of squares for the full and reduced models with $r$ and $s$ parameters, respectively. Using this test, $y$ is declared Granger causal for $x$ if the observed test statistic $F$ exceeds the $(1-\alpha)\%$ quantile of an F-distribution with $r-s$ and $T-r$ degrees of freedom. Alternatively, one can also use a $\chi^2$ statistic based on likelihood ratio or Wald statistics \citep{cromwell1994}. 
A key step in carrying out the testing is to identify the model's order (or lag), $d$. 
We further discuss lag selection in Section~\ref{sec:penalizedngc}. Alternatively, one can also use tests in the spectral domain, using Fourier or wavelet representations \citep{geweke1982, dhamala2008}.

\begin{figure}[t]
    \centering
    \includegraphics[page=1,width=0.75\textwidth, clip=TRUE, trim=3cm 8cm 4cm 3cm]{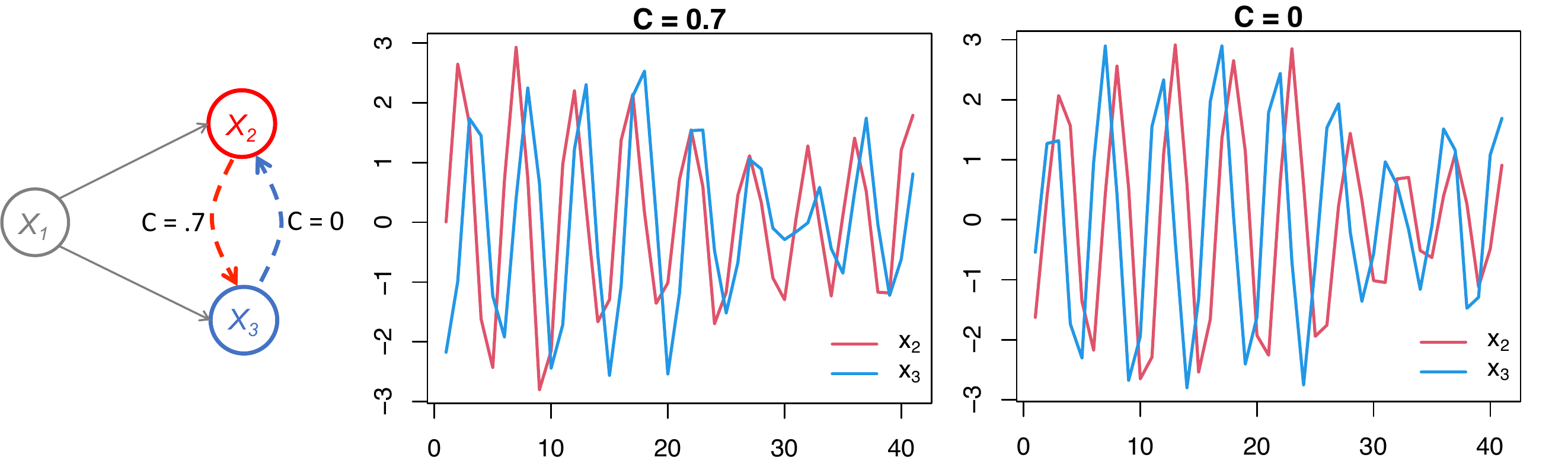}    
    \caption{A simple VAR process with three variables generated according to Eq.(\ref{eqn:3varex}). The time series plots (right) suggest Granger causal interactions between $x_2$ and $x_3$ in a bivariate analysis excluding $x_1$. Moreover, the direction of causality is different when $C = 0.7$ ($x_2 \to x_3$) and $C=0$ ($x_3 \to x_2$). Bivariate VAR modeling using the \texttt{vars} R-package  \citep{varspackage} confirms these observations.}
    \label{fig:threevarex}
\end{figure}

Regardless of testing procedure, Granger causality based on only two variables severely limits the interpretation of the findings: without adjusting for all relevant covariates, a key assumption of Granger causality is violated. This limitation, which has been well documented \citep[see, e.g.][]{lutkepohl1982}, is illustrated in Figure~\ref{fig:threevarex}. 
Here, data is generated according to the following simple VAR process with three variables and i.i.d. innovations $e_{t,i} \sim N(0,0.1^2)$:
\begin{align}\label{eqn:3varex}
    x_{t,1} &= .5 x_{t-1,1} -.8 x_{t-2,1} + e_{t,1} \nonumber \\ 
    x_{t,2} &= .5 x_{t-1,2} -.8 x_{t-2,2} + C x_{t-1,1} + (.7-C) x_{t-2,1} +  e_{t,2} \\ 
    x_{t,3} &= .5 x_{t-1,3} -.8 x_{t-2,3} + (.7-C) x_{t-1,1} + C x_{t-2,1} +  e_{t,3}\,. \nonumber 
\end{align}
The two time series plots in Figure~\ref{fig:threevarex} correspond to two different VAR models: 
one with $C = 0.7$ and another with $C = 0$. In the first model, $x_2$ and $x_3$ are affected by values of $x_1$ in lags 1 and 2, respectively. This relationship is reversed in the second model. The clear patterns of $x_2$ and $x_3$ in the time series plots in Figure~\ref{fig:threevarex} suggest that ignoring $x_1$, we may either conclude that $x_2$ is Granger causal for $x_3$ (when $C = 0.7$) or that $x_3$ is Granger causal for $x_2$ (when $C = 0.7$). This observation is indeed confirmed when we use a test of Granger causality in either case, highlighting the limitation of bivariate tests of Granger causality. 

In spite of their limitations, bivariate tests of Granger causality have been widely used in many application areas, from economics \citep{chiou2008economic} and finance \citep{hong2009finance} to neuroscience \citep{seth2015neuro} and meteorology \citep{mosedale2006ocean}. In the next section, we discuss recent developments that aim to mitigate this limitation by analyzing a potentially large set of variables.

\section{NETWORK GRANGER CAUSALITY}\label{sec:ngc}
The limitations of identifying Granger causality using bivariate models---illustrated in the three-variable example of Fig.~\ref{fig:threevarex}---have long been known and discussed in the literature~\citep[e.g.,][]{sims1980}. 
Needing to account for many variables when identifying Granger causality arises in at least two settings. First, when the goal is to investigate Granger causality between two (or a handful of) \emph{endogenous} variables $x$ and $y$, we need to account for the remaining \emph{exogenous} variables ---targeting the notion of ``\emph{all} other relevant information''---to prevent identifying incorrect Granger causal relations. This is the setting illustrated in Fig.~\ref{fig:threevarex}, and is common in macroeconomic and econometric studies \citep{bernanke2005explains}. 
Methods based on \emph{summaries} of exogenous variables, using, e.g., latent factors, have been commonly used to achieve this goal \citep{bernanke2005}. 

In the second setting, which arises naturally in the study of many physical, biological and social systems, the goal is to investigate the relationships among all the variables from a systems perspective. In this case, \emph{all variables} are endogenous. 
For instance, when learning gene regulatory networks, all the genes in a given biological pathway are of interest. Similarly, when studying brain connectivity networks, the goal is to interrogate interactions among all regions of interests (ROIs) in the brain. These applications have led to the development of methods for identifying Granger causal relationships among a large set of variables, which can be compactly represented as a network or graph \citep{eichler2012graphical} (see Fig.~\ref{fig:ngcbasics}) and underlie the study of \emph{network Granger causality} \citep{basu2015ngc}.

\subsection{Granger causality based on VARs}\label{sec:earlyngc}
In this section we explicitly consider the popular VAR model for Granger causality analysis of multiple variables:
\begin{equation}\label{eqn:var}
    \x_t = \sum_{k=1}^d A^k \x_{t-k} + \e_t,
\end{equation}
where variables and parameters are defined as in (\ref{eqn:linmodel}). 

\begin{figure}[t]
    \centering
    \includegraphics[page=4,width=0.95\textwidth, clip=TRUE, trim=1.5cm 7cm 2cm 4cm]{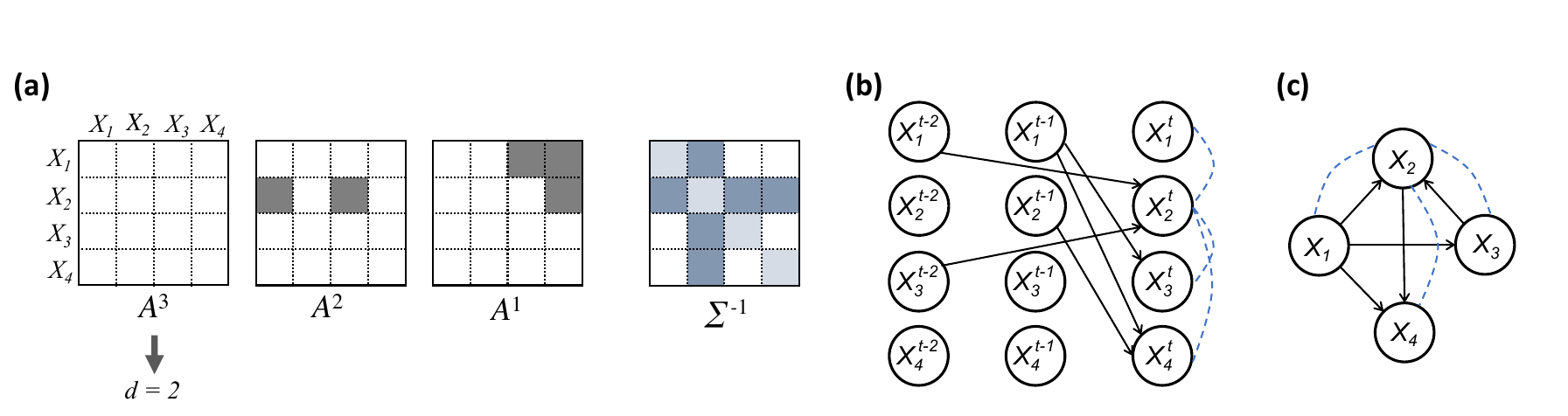}    
    \caption{Illustration of link between network Granger causality and parameters of SVAR models. (a) Lag matrices $A^1, \ldots, A^d$ and inverse covariance matrix of the innovation, $\Sigma^{-1}$, of an SVAR model. Nonzero entries of $A^k$ and $\Sigma^{-1}$ are shaded. 
    (b) \emph{Expanded} graphical model, which replicates variables over time. (c) \emph{Compact} graphical model combining all interactions from past lags. In both graphs, Granger causal interactions (solid edges) correspond to nonzero entries in $A^k$ and instantaneous causal effects (dashed undirected edges) nonzero entries in $\Sigma^{-1}$.}
    \label{fig:ngcbasics}
\end{figure}

\begin{proposition} \label{prop:gcdef} Straightforwardly following from the bivariate case \citep{granger1969},  series $x_i$ is Granger causal for series $x_j$ if and only if $A^k_{ij} \ne 0$ for some $1 \le k \le d$.
\end{proposition}

Reading off statements of Granger non-causality from the zeros of the lag matrices is illustrated in Fig.~\ref{fig:ngcbasics}. The Granger causal relations can also be described via two different graphical models \citep{eichler2012graphical}: The first is an \emph{expanded} graph (Fig.~\ref{fig:ngcbasics}b) with $p$ nodes for each time point $t, t-1, \ldots, t-d$ and edges corresponding to nonzero entries in $A^k$. This representation is similar to that in dynamic Bayesian networks 
\citep{ghahramani1997learning}. 
The second graph is a \emph{compact} representation (Fig.~\ref{fig:ngcbasics}c), combining edges from different lags of the expanded graph. This latter graph captures the Granger causal relations. In addition, undirected edges indicate instantaneous dependencies captured by nonzero entries in the inverse covariance matrix $\Sigma^{-1}$ of the innovations, $\e_t$. 

Despite the direct connection between Granger causality and nonzero entries of $A^k$ (Proposition~\ref{prop:gcdef}), earlier VAR-based approaches used tests of variance similar to those for bivariate models in (\ref{eqn:Ftest}). Moreover, concerned with the increasing number of parameters in the model---$O(p^2)$ parameters for a model with $p$ variables---earlier approaches focused on few time series. \citet{bernanke2005} state that ``To conserve degrees of freedom, standard VARs rarely employ more than six to eight variables.''
While a step forward, it is difficult to argue that early \emph{moderate-dimensional} approaches account for all the relevant information when determining Granger causal relations. Thus, these approaches still do not satisfy the requirements of definition (\ref{eqn:gcdef}). 
This limitation was underscored by \citet{bernanke2005} when stating that ``[The] small number of variables is unlikely to span the information sets used by actual central banks [...].'' We consider the challenge of scaling to a large number of series under the two scenarios outlined above: assuming a large set of \emph{exogenous} series, or that all series are \emph{endogenous}.

To account for a number of exogenous variables when studying the relationship between a small number of endogenous variables, a well-known approach is the factor-augmented VAR (FAVAR) model of \citet{bernanke2005}: 
\begin{equation}\label{eqn:favar}
    \begin{pmatrix}
    \x_t \\ \f_t
    \end{pmatrix} = 
    \sum_{k=1}^d \widetilde A^k 
    \begin{pmatrix}
    \x_{t-k} \\ \f_{t-k}
    \end{pmatrix}
    + \widetilde\e_t.
\end{equation}
This model is seemingly similar to the VAR model in (\ref{eqn:var}). However, the $m$-dimensional factors $\f_t$---representing exogenous variables---are \emph{unobserved}. 
\citet{bernanke2005} propose two estimation procedures for (\ref{eqn:favar}) with constraints on the factors: a two-step procedure based on principal components, and a direct estimation procedure based on maximum likelihood.
Factor models have been used extensively in econometrics \citep{stock2011dynamic}. Followup work has further investigated the estimability of the parameters \citep{belviso2006structural} and the choice of number of unobserved factors \citep{ahn2013eigenvalue, onatski2010determining, amengual2007consistent}. 

\begin{figure}
    \centering
    \includegraphics[page=2,width=0.8\textwidth, clip=TRUE, trim=2cm 7cm 2cm 2cm]{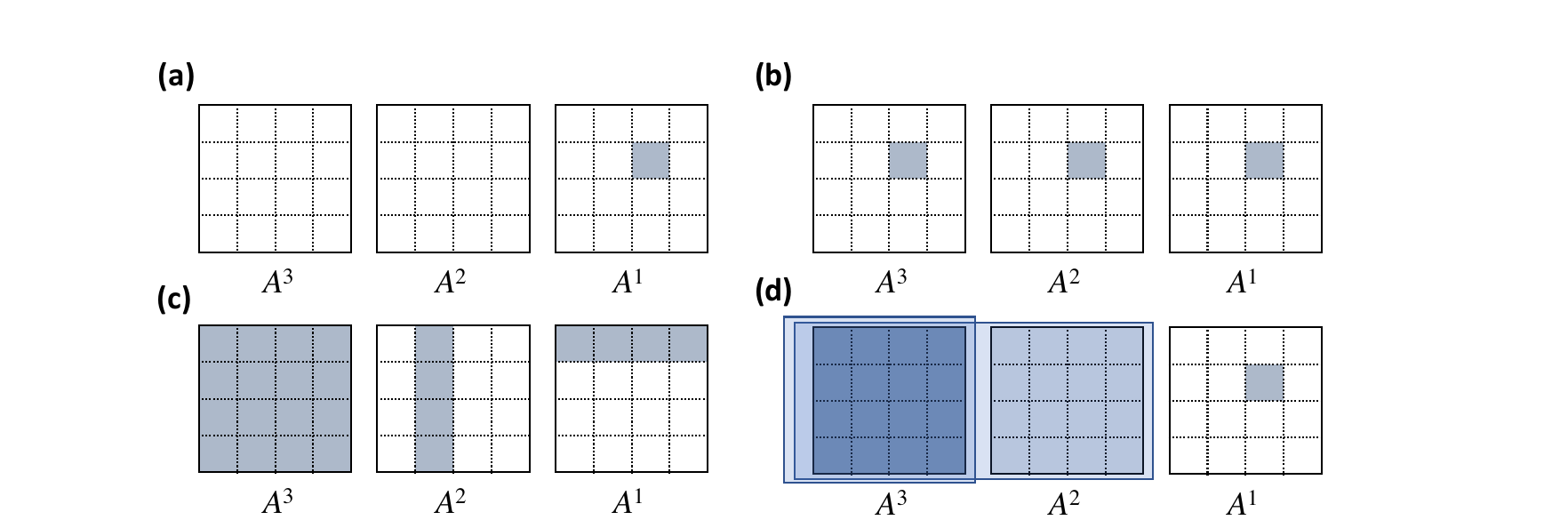}    
    \caption{Illustration of different sparsity-inducing penalties for Granger causality estimation based on VARs: (a) the lasso penalty $|A^k_{ij}|$ applied to each entry of lag matrices \citep{fujita2007}; (b) the group lasso penalty $\| (A^1_{ij}, A^2_{ij} \ldots, A^d_{ij}) \|_2$ applied to all lags of the same entry $(i,j)$ \citep{lozano2009}; (c) general group lasso penalty \citep{basu2015ngc}, applied to groups of related variables or entire lag matrices $A^k$; (d) joint lasso and hierarchical group lasso penalties for inducing sparsity while selecting lags by forcing $A^k=0$ for larger $k$ \citep{nicholson2017varx}.}
    \label{fig:penalizedngc}
\end{figure}

The second scenario involves fitting VAR models with a large number of endogenous variables. 
Earlier approaches primarily used shrinkage penalties to obtain reasonable estimates in moderate-dimensional VAR models, followed by classical test-based approaches (e.g., the F-test) to infer Granger causality. 
For instance, motivated by earlier work \citep{litterman1986}, \citet{leeper1996} considered a Bayesian approach using a prior shrinking large coefficients or distant lags. 
Recent work has increasingly focused on directly selecting the nonzero entries of the $A^k$s via sparsity-inducing penalties, often by augmenting the VAR loss function. 
For the commonly used least squares loss and a general penalty $\Omega(\cdot)$ on the coefficient matrices $A^1,\ldots,A^d$, the general problem can be written as 
\begin{equation}\label{eqn:penalizedls}
    \min _{A^1,
    \ldots,A^d \in \mathbb{R}^{p\times p}} \sum_{t=d+1}^{T}\left\|\x_{t}- \sum_{k=1}^{d}A^k \x_{t-k}  \right\|_2^2 + \Omega\left(A^1, \ldots, A^d\right), 
\end{equation}
where $\|\cdot\|_2$ denotes the $\ell_2$ norm and $T$ the length of the time series. \citet{fujita2007} proposed to estimate \emph{high-dimensional} VARs by using a lasso penalty \citep{tibshirani1996}: 
\[
\Omega(A^1,\ldots,A^d) = \lambda \sum_{k=1}^d\sum_{i,j=1}^p|A^k_{ij}|,
\]
with $\lambda$ a \emph{tuning parameter} controlling \emph{element-wise sparsity} in $A^k$, encouraging many entries to be exactly zero. One can directly deduce from the lasso estimate that \emph{$x_i$ is Granger causal for $x_j$ if there exists $1 \le k \le d$ such that $A^k_{ij} \ne 0$}. See Fig.~\ref{fig:penalizedngc}a.
The motivating application for \citet{fujita2007} was the estimation of gene regulatory networks; based on the particulars of this application, they developed their method for \emph{panel data}, which often contain observations over a small number of time points, but with repeated measures for multiple subjects. \citet{chudik2011} considered a very similar estimator (using also a lasso penalty) for economic time series data.

\begin{figure}[t]
    \centering
    \includegraphics[page=6,width=0.9\textwidth, clip=TRUE, trim=0cm 6.5cm 0cm 2cm]{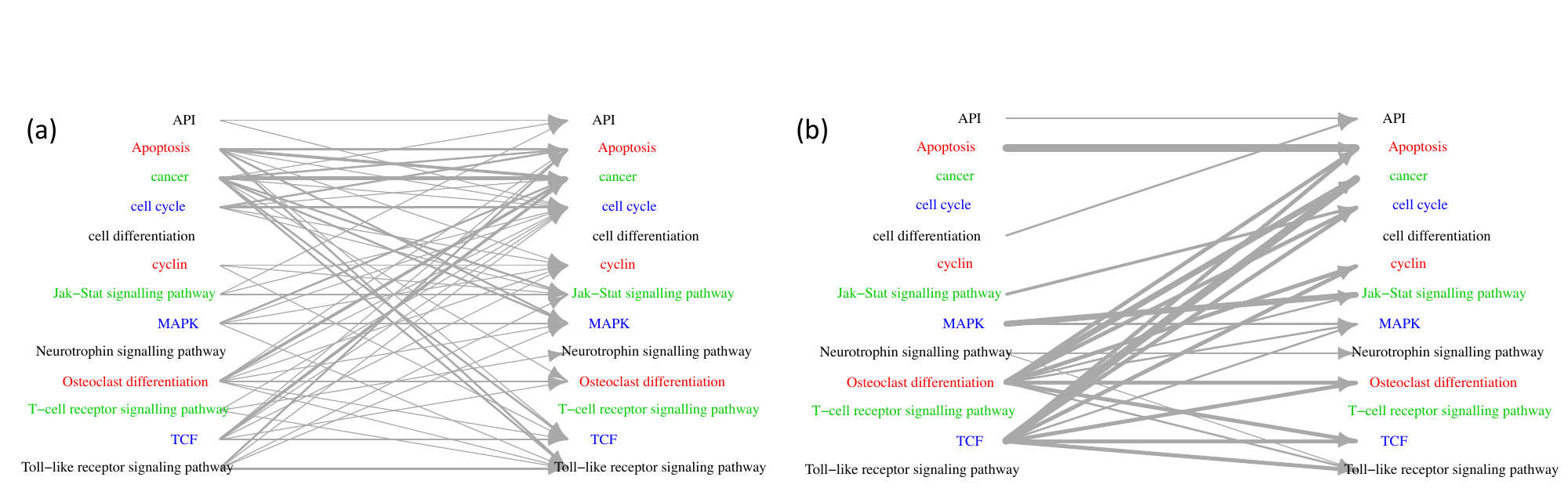}    
    \caption{(a) Lasso vs. (b) group lasso estimates of T-cell gene regulatory networks \citep{basu2015ngc}.}
    \label{fig:lassovsglasso}
\end{figure}

\citet{lozano2009} used a group lasso penalty \citep{yuan2006} for aided Granger causality interpretability: 
$$
\Omega(A^1,\ldots,A^d) = \lambda \sum_{i,j=1}^p
\left\|(A^1_{ij}, \ldots, A^d_{ij})\right\|_2.
$$ 
This penalty, which is depicted in Fig.~\ref{fig:penalizedngc}b,  corresponds directly to Granger non-causality from $x_i$ to $x_j$ by enforcing $A^k_{ji} = 0$ for all $k$. \citet{basu2015ngc} considered more general group lasso penalties, to not only group over lags, but also sets of related variables and even entire matrices (see Fig.~\ref{fig:penalizedngc}c). 
The authors also showed that the sparsity pattern resulting from group lasso penalty is only consistent with the truth if the grouped coefficients have similar magnitudes, and that group lasso may only achieve \emph{directional consistency}; they proposed a thresholded group lasso penalty to consistently learn the sparsity patterns. 
As illustrated in Fig.~\ref{fig:lassovsglasso}, the resulting estimates can facilitate the interpretation of Granger causal effects in settings with many variables. 

The general estimation framework in (\ref{eqn:penalizedls}) has been extended to  account for dependencies in the inverse covariance of the innovations, $\Sigma^{-1}$ \citep{davis2016}, and to to combine the ideas of sparsity and unobserved exogenous variables \citep{basu2019low}. 
Asymptotic properties of the resulting estimators have also been investigated in high-dimensional settings, where $p \gg T$ \citep{bickel2011, basu2015}. In particular, \citet{basu2015} established a connection between the sample size ($T$) needed for high-dimensional consistency of the lasso estimate of a VAR process and the eigen-structure of its spectral density matrix. 
More recent work has developed asymptotically valid inference for the estimated parameters of the VAR process \citep{neykov2018unified, zheng2019testing, zhu2020confidence}. 
Some of these developments have also been implemented in publicly available software packages, including \texttt{mgm} \citep{mgmpackage}, \texttt{bigvar} \citep{bigvarpackage} and \texttt{ngc} \citep{ngcpackage}. 

Bayesian approaches have also been considered as an alternative to regularization methods for analyzing large VAR processes. For instance,  \citet{george2008bayesian} proposed a Bayesian stochastic search algorithm to identify high-dimensional VAR processes, whereas \citet{banbura2010large} showed that better performance can be achieved in large models if the tightness of the priors is increased as the model size increases.
More recently, \citet{ahelegbey2016sparse} considered sparsity inducing priors for high-dimensional VAR processes, \citet{ghosh2019high} established posterior consistency of the Bayesian estimates when using sparsity inducing priors, and \citet{billio2019bayesian}  proposed nonparametric Bayesian priors that cluster the VAR coefficients and induce group-level shrinkage.

\subsection{Lag Selection and Non-Stationary VAR Models}\label{sec:penalizedngc}

In classical linear VAR methods, one must explicitly specify the maximum time lag, $d$, when assessing Granger causality. 
Early approaches often set $d$ based on prior knowledge or in \emph{ad hoc} ways. VARs with different lags may result in different conclusions, further complicating the interpretation of Granger causality. 
If the specified lag is too short, Granger causal connections at longer lags will be missed, while overfitting may occur if the lag is too large, a problem exacerbated by high-dimensional VAR models. 

\begin{figure}
    \centering
    \includegraphics[page=5,width=0.9\textwidth, clip=TRUE, trim=1cm 9cm 1cm 3cm]{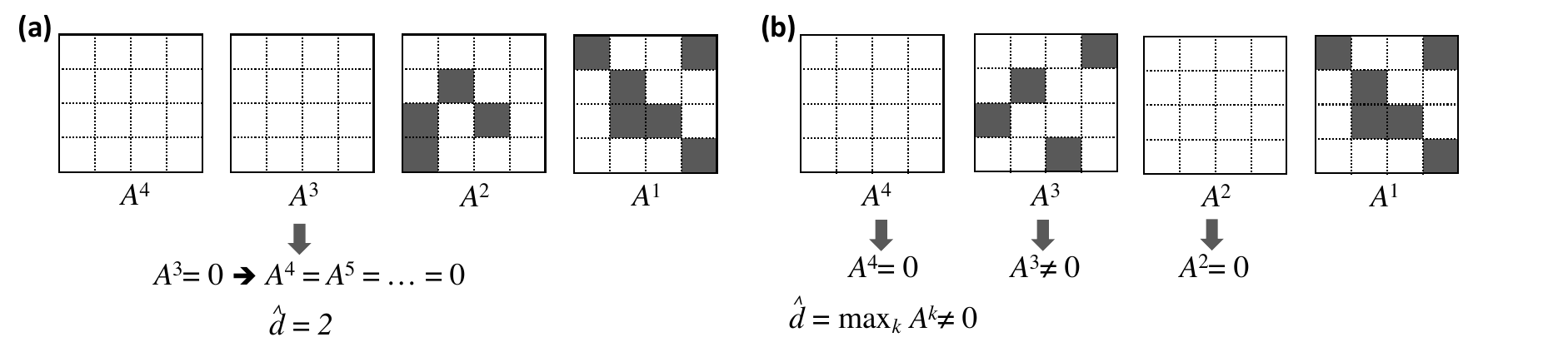}    
    \caption{Illustration of two approaches for lag selection: (a) Assuming a \emph{decay assumption}---that is, $A^k = 0 \Rightarrow A^{k'} = 0 \, \forall \, k' \ge k$---the lag $d$ can be estimated by identifying the first $k$ such that $A^k = 0$; (b) The lag $d$ can be estimated without assuming a decay assumption by enforcing entire lag matrices to be zero and setting $\hat d = \max_k \hat A^k \ne 0$.}
    \label{fig:lagselection}
\end{figure}

Regularization-based approaches can be used to systematically estimate the optimal lag $d$ from data. To this end, \citet{shojaie2010} proposed a \emph{truncating lasso} penalty that shrinks entire coefficient matrices $A^k$ to zero and then sets all following $A^{k+1}$ to zero (see Fig.~\ref{fig:lagselection}a). The idea is to scale the penalty for each $A^k$ using data-driven weights calculated based on previous coefficient matrices in previous lags $A^{k-1}$. Formally, the penalty is given by:
\[
    \Omega(A^1, \ldots, A^T) = \lambda\sum_{k=1}^T \omega^k {\sum_{i,j=1}^p{|A^k_{ij}|}},
\]
where $\omega^1 = 1$ and for $k \ge 2$ the weights can be compactly written as
\[    
    \omega^k = 
    \mathbb{I}\left(A^{k-1}; \Big\{A: (T - k) \left\|A\right\|_0 \ge p^2 \beta\Big\} \right),
\]
with $\mathbb{I}(A; E) = 0$ if $A \in E$ and $\mathbb{I}(A;E) = \infty$ if $A\notin E$ (the convex indicator function). Here, $\|A\|_0$ gives the number of nonzero entries of $A$ and $\beta$ is a second tuning parameter. \citet{shojaie2010} show that a block-coordinate descent algorithm convergences to a local minima and establish consistency of this algorithm for selecting the correct Granger causality network in high-dimensional panel data settings. They also propose error-based choices for the two tuning parameters ($\lambda$ and $\beta$) that control the type-I and II errors in selecting Granger causal effects. 

While the \emph{decay assumption} of \citet{shojaie2010} may be satisfied in some applications, it may fail in others. To overcome this limitation, \citet{shojaie2012} proposed an adaptive thresholded lasso penalty that can data-adaptively set entire lag matrices to zero, while allowing others to be nonzero. The effect of this penalty, depicted in Fig.~\ref{fig:lagselection}b, is somewhat similar to the effect of the \emph{automatic relevance determination} (ARD) priors proposed in the Bayesian nonparametric approach of \citet{fox2011bayesian} for switching dynamic linear models. More specifically, the ARD prior turns off entire blocks of $A^k$ based on the value of their corresponding precision parameters. Another approach for automatic lag selection using regularization, proposed by \citet{nicholson2017varx} is to use a \emph{hierarchical group lasso} penalty, depicted in Fig.~\ref{fig:penalizedngc}d. The hierarchical penalty is based on a decay assumption, similar to that in \citet{shojaie2010}, but is convex and can thus lead to more computationally efficient implementation.

The Bayesian nonparametric approach of \citet{fox2011bayesian} addresses another limitation of classical Granger causality methods based on VARs: the assumption of \emph{stationary}. \citet{fox2011bayesian} relax this assumption by considering a \emph{switching} VAR model, with lag matrices $A^k$ a function of a latent (switching) variable $z_t$; in other words, $A_t^k = A^k(z_t)$, where the distribution of $z_t$ depends on $z_{t-1}$. \citet{fox2011bayesian} also consider a switching \emph{state-space} model allowing the observed data to be a noisy version of the switching VAR process. \citet{nakajimawest2013} instead propose a method for inducing \emph{continuously-varying} (rather than switching) sparsity in a time-varying VAR (TV-VAR) model through the use of a latent threshold process.  A vectorized form of the time-varying lag matrices is assumed to follow a VAR(1) process with elements thresholded to zero based on a set of latent threshold variables.  \citet{nakajimawest2013} consider a Bayesian approach to inference in this model.

An alternative approach for handling non-stationarity was recently proposed by \citet{safikhani2020} in the setting of high-dimensional piece-wise VAR processes with many structural break points. To consistently identify the break points and learn the coefficient parameters in each regime, the authors consider a reparameterization based on changes in lag matrices,  $\Delta^t = A^t - A^{t-1}$, and use a combination of lasso penalized estimation and model selection based on Bayesian information criterion (BIC) to enforce piece-wise stationarity in estimated lag matrices. \citet{bai2020multiple} have recently used similar ideas in the case where the lag matrices are a combination of sparse and  low-rank components, capturing non-stationary VAR models in the presence of (unobserved) exogenous variables.

\section{MORE GENERAL NOTIONS OF GRANGER CAUSALITY}\label{sec:general}

The notion of Granger causality explored so far is suitable for time series that follow linear dynamics. However, many interactions in real-world applications, like neuroscience and genomics, are inherently nonlinear. In these cases, using linear models may lead to inconsistent estimation of Granger causal interactions. Furthermore, classical Granger causality analyses assume real-valued Gaussian time series.  This restriction has hindered Granger causality analysis of many important applications involving, for example, count or categorical time series. 

To generalize the VAR model of (\ref{eqn:var}), consider a process that \emph{component-wise} can be written as follows:
\begin{align}
\label{eq:compVAR}
x_{t i}=g_{i}\left(x_{<t 1}, \ldots, x_{<t p}\right)+e_{t i}.
\end{align}
Here, $g_i$ is a function specifying how the past of all $p$ series map to a particular series $i$. Assuming diagonal $\Sigma$, the linear VAR model is a special case of (\ref{eq:compVAR}) with $g_{i}$ a linear function with coefficients given by the $i$th row of coefficient matrices, $A^k$. 
In contrast to standard multivariate forecasting, where a function $g$ would jointly model all outputs $\x_t$, this component-wise specification is more immediately amenable to Granger causal analysis. In particular, we can extend the definition of Granger causality to this more expressive class of dynamical models by noting that if the function $g_i$ does not depend on $x_{<tj}$, then $x_j$ is irrelevant in the prediction of series $x_i$.  
Formally,

\begin{definition} \label{def:granger_g}
Time series $x_j$ is Granger non-causal for time series
$x_i$ iff for all $\left(x_{<t 1}, \ldots, x_{<t p}\right)$ and all $x^{\prime}_{<tj}\neq x_{<t j}$,
\begin{align}
g_{i}\left(x_{<t 1}, \ldots, x_{<t j}, \ldots, x_{<t p}\right) &= g_{i}\left(x_{<t 1}, \ldots, x^{\prime}_{<t j}, \ldots x_{<t p}\right); \nonumber
\end{align}
that is, $g_i$ is invariant to $x_{<tj}$.
\end{definition}

Related definitions for specific classes of models have appeared in the literature~\citep[see e.g.,][]{eichler2012graphical}.  Note that (\ref{eq:compVAR}) still assumes additive noise. Definition~\ref{def:granger_g} can be further generalized to statements of conditional independencies modeling arbitrary non-linear relationships between time series, referred to as \emph{strong Granger causality} \citep[e.g.,][]{Florens:1982}. 
Building on the \emph{component-wise} process of (\ref{eq:compVAR}), we further define Granger causality in situations where the series at time $t$ are conditionally independent of one another given the past realizations:
\begin{equation}
\label{eq:fact}
p\left(\x_{t} | \x_{<t}\right)=\prod_{i=1}^{p} p\left(x_{ti} | \x_{<t}\right).
\end{equation}

\begin{definition}\label{def:strongGranger}
Time series $x_j$ is Granger non-causal for time series $x_i$ iff $\forall t$,
\begin{align}
p\left(x_{i t} | x_{<t1}, \ldots, x_{<tj}, \ldots, x_{<tp}\right)= 
p\left(x_{i t} | x_{<t1}, \ldots, x_{<t(j-1)}, x_{<t(j+1)}, \ldots, x_{<tp}\right).
\end{align} 
\end{definition}

In the context of these more general notions of Granger causality, we review in Sections~\ref{sec:categorical}-\ref{sec:nonlinear} recent advances for analyzing multivariate discrete-valued and non-linear time series, as well as multivariate point processes. 

Another implicit assumption of classical Granger causality is that the time series of interest are observed at a regular sampling rate that matches the causal scale. However, due to data integration across heterogeneous sources, many data sets in econometrics, health care, environment monitoring, and neuroscience comprise multiple series sampled at different rates, referred to as \emph{mixed-frequency} time series. Furthermore, due to the cost or data collection challenges, many series may be sampled at a rate lower than the true causal scale of the underlying process. For example, many econometric indicators, such as GDP and housing price data, are recorded at quarterly and monthly scales \citep{moauro:2005}, but important interactions between these indicators may occur weekly or bi-weekly \citep{moauro:2005,stram:1986,boot:1967}. In neuroscience, imaging modalities with high spatial resolution, like fMRI, 
have relatively low temporal resolutions, but many important neuronal processes and interactions happen at finer time scales \citep{zhou:2014}. A causal analysis at a slower time scale than the true causal time scale may miss true interactions and add spurious ones \citep{zhou:2014,silvestrini:2008,boot:1967,breitung:2002}. In Section~\ref{sec:mixedFreqGranger}, we review recent approaches to identifying Granger causality in \emph{subsampled} and \emph{mixed frequency} time series~\citep{Gong:2015,tank2019biometrika}.

\subsection{Discrete-valued time series}
\label{sec:categorical}

A variety of applications give rise to multivariate discrete-valued time series, including count, binary, and categorical data.  Examples include voting records of politicians, discrete health states for a patient over time, or action labels for players on a team.  Furthermore, even when the raw recording mechanism produces continuous-valued time series,
to facilitate downstream analyses, the series may be quantized into a small set of discrete values; examples include weather data from multiple stations \citep{Doshi-Velez2011Infinite}, wind data \citep{Raftery1985Model}, stock returns \citep{nicolau2014new}, or sales volume for a collection of products \citep{Ching2002MultivariateMarkov}.  In these cases, the traditional VAR framework for Granger causal analysis, (\ref{eqn:var}), is inappropriate.  In this section, we review recently proposed models, based on the more general framework of Definitions~\ref{def:granger_g} and \ref{def:strongGranger}, that infer Granger causality using multivariate, discrete-valued time series.

\subsubsection{Categorical time series}

\begin{figure}[t]
\includegraphics[page=9,width=0.95\textwidth, clip=TRUE, trim=0cm 8.8cm 0cm 4cm]{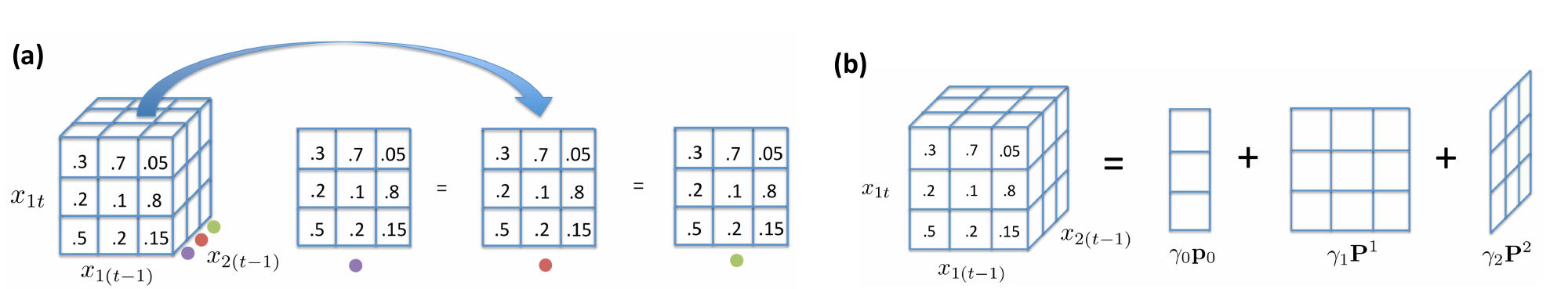}    
\caption{(\emph{a}) Illustration of Granger non-causality in an example with $p = 2$ and $m_1 = m_2 = 3$.
Since the tensor represents conditional probabilities, the columns of the front face of
the tensor, the vertical $x_{t1}$ axis, must sum to one. Here, $x_2$ is not Granger causal for
$x_1$ since each slice of the conditional probability tensor along the $x_2$ mode is equal. (\emph{b}) Schematic of the MTD factorization of the conditional probability tensor
$p\left(x_{t1} | x_{(t-1)1}, x_{(t-1)2}\right)$. 
\label{fig:transitionTensor}}
\end{figure}

Consider a multivariate categorical time series $\x_t$, and let $m_{i}$ represent the number of categories that series $i$ may take. An order \(k\) multivariate Markov chain models the transition probability between
the categories at lagged times \(t-1, \ldots, t-k\) and those at time \(t\) using a transition probability distribution; under the simplifying assumption of (\ref{eq:fact}):
\begin{equation}
\label{eq:2.2}
p\left(\x_{t} | \x_{<t}\right)=\prod_{i=1}^{p} p\left(x_{ti} | \x_{t-1}, \ldots, \x_{t-k}\right).
\end{equation}
The component-wise structure of the assumed transition distribution enables estimation and inference to be divided into independent subproblems over each series, $x_i$.  Additionally, Granger non-causality follows Definition~\ref{def:strongGranger}: Analyzing the \emph{transition probability tensor} for $p\left(x_{ti} | \x_{t-1}, \ldots, \x_{t-k}\right)$, $x_j$ does not Granger cause $x_i$ if all subtensors along the mode associated with $x_j$ are equal.  See Fig.~\ref{fig:transitionTensor}. 

Unfortunately, discovering such invariances (equivalence amongst subtensors) via, e.g., penalized likelihood proves computationally prohibitive in even moderate dimensions.  Instead, \citet{tank2021categorical} propose a more tractable yet still flexible parameterization of the transition probabilities leveraging the mixture transition distribution (MTD)~\citep{berchtold2002Mixture,Raftery1985Model}: 
\begin{equation}
\label{eq:2.5}
p\left(x_{ti} | x_{(t-1)1}, \ldots, x_{(t-1)p}\right)=\gamma_{0} p_{0}\left(x_{ti}\right)+\sum_{j=1}^{p} \gamma_{j} p_{j}\left(x_{ti} | x_{(t-1)j}\right),
\end{equation}
where $p_0$ is a probability vector, $p_{j}( \cdot | \cdot)$ is a pairwise transition probability table between $x_{(t-1)j}$ and $x_{ti}$ and $\gamma=\left(\gamma_{0}, \gamma_{1}, \ldots, \gamma_{p}\right)$ is a $p + 1$ dimensional probability distribution
such that $\mathbf{1}^{T} \gamma=1$ with $\gamma_{j} \geq 0$, $j=0, \ldots, p$. \citet{tank2021categorical} show that the intercept term, $p_0$, which is not traditionally included in MTD models, is critical for model identifiability, and thus, Granger causality. The framework of \citet{tank2021categorical} is general for higher-order lags and $t-1$ is presented here for ease of exposition.  Additionally, interaction terms can also be included in the MTD decomposition.  See Fig.~\ref{fig:transitionTensor} for a visualization of the MTD transition probability tensor decomposition.

The MTD model---originally proposed for parsimonious modeling of higher order Markov chains---has been plagued by a \emph{non-convex objective} and \emph{unknown identifiability conditions} that have limited its utility~\citep{nicolau2014new,Zhu2010NewEstimation,Berchtold2001Estimation}.
\citet{tank2021categorical} instead propose a change-of-variables reparameterization of the MTD that straightforwardly addresses both issues, thus enabling practical application of the MTD model to Granger causality selection.
Let $\mathbf{p}^{0}$ denote the vector of intercept probabilities, $\mathbf{p}_{x_{i t}}^{0}=p_{0}\left(x_{i t}\right)$, and $\mathbf{P}^j \in \mathbb{R}^{m_{i} \times m_{j}}$ the pairwise transition probability matrix
$\mathbf{P}^j_{x_{ti}, x_{(t-1)j}}=p_{j}\left(x_{ti} | x_{(t-1)j}\right)$.  Let $\mathbf{Z}^{j}=\gamma_{j} \mathbf{P}^{j}$ and $\mathbf{z}^{0}=\gamma_{0} \mathbf{p}^{0}$. Then, the factorization of the conditional probability tensor for the MTD in (\ref{eq:2.5}) can be rewritten as
\begin{equation}
\label{eq:3.2}
p\left(x_{ti} | x_{(t-1)1}, \ldots, x_{(t-1)p}\right)=\mathbf{z}_{x_{ti}}^{0}+\sum_{j=1}^{p} \mathbf{Z}_{x_{ti} x_{(t-1j)}}^{j}.
\end{equation}
\begin{proposition}\label{Proposition 3.3}\citep{tank2021categorical}
In the MTD model of (\ref{eq:3.2}), following Definition~\ref{def:strongGranger}, time series $x_j$ is Granger non-causal for time series $x_i$ if and only if the columns of $\mathbf{Z}^{j}$ are all equal. Furthermore, all equivalent MTD model parameterizations give the same Granger causality conclusions.
\end{proposition} 
Intuitively, if all columns of $\mathbf{Z}^{j}$ are equal, the transition distribution for $x_{ti}$ does not depend on $x_{(t-1)j}$.
This result for MTD models is analogous to the general Granger non-causality result for the slices of the conditional probability tensor being constant along the $x_{(t-1)j}$ mode being equal. 

The optimization problem for maximizing log-likelihood can be written as follows.  Letting
\begin{equation}
L_{\mathrm{MTD}}(\mathbf{Z})=-\sum_{t=1}^{T} \log \left(\mathbf{z}_{x_{i t}}^{0}+\sum_{j=1}^{p} \mathbf{Z}_{x_{ti}x_{(t-1)j}}^{j}\right),
\end{equation}
and including the necessary probability constraints (positivity and summing to one), we have:
\begin{equation}
\label{eq:3.3}
\begin{aligned}
&\underset{\mathbf{Z}, \gamma}{\operatorname{minimize}} \,\, L_{\mathrm{MTD}}(\mathbf{Z}) \\
&\text { subject to } \mathbf{1}^{T} \mathbf{Z}^{j}=\gamma_{j} \mathbf{1}^{T}, \mathbf{Z}^{j} \geq 0, \forall j \quad \mathbf{1}^{T} \gamma=1, \gamma \geq 0.
\end{aligned}
\end{equation}
Problem (\ref{eq:3.3} is convex since the objective function is a linear function composed with a log function and only involves linear equality and inequality constraints \citep{boyd2004convex}.

\begin{figure}[t]
\centering
\includegraphics[page=7,width=0.75\textwidth, clip=TRUE, trim=5cm 8.5cm 6cm 7cm]{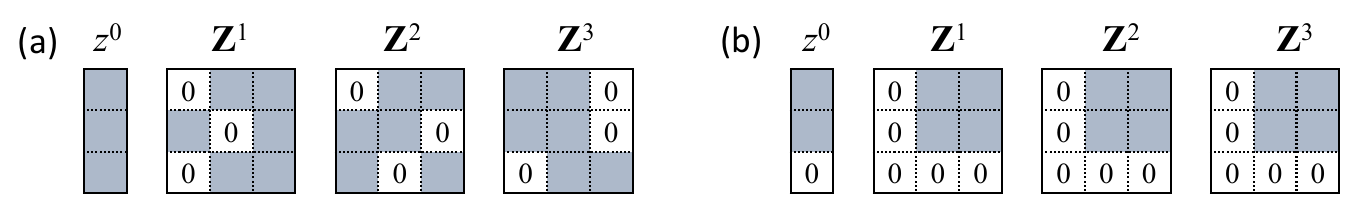}    
\caption{Schematic of identifiability conditions for the (\emph{a}) MTD and (\emph{b}) mLTD with $d = 3$ and $m_1 = m_2 = m_3 = 3$. Identifiability
for MTD requires a zero entry in each row of $\mathbf{Z}^{j}$; for mLTD the first column
and last row must all be zero. In MTD, the columns of each $\mathbf{Z}^{j}$ must sum to the
same value, and must sum to one across all $\mathbf{Z}^{j}$. \label{fig:MTDmLTDidentifiability}}
\end{figure}

The $\mathbf{Z}^{j}$ reparameterization in (\ref{eq:3.2}) provides clear intuition for why the MTD model may not be identifiable. Since the probability function is a linear sum
of $\mathbf{Z}^{j}$s, one may take mass from some $\mathbf{Z}^{j}$ and move it to some $\mathbf{Z}^{k}$, $k \neq j$ or $\mathbf{z}^0$, while keeping the conditional probability tensor constant. 
These sets of equivalent MTD parameterizations---that yield the same factorized conditional distributions---$p\left(x_{ti} | \x_{(t-1)}\right)$ forms a convex set~\citep{tank2021categorical}. Taken together, the convex reparameterization and this result implies that the convex function given in (\ref{eq:3.3}) has no local optima, and that the globally optimal solution is given by a convex set of equivalent MTD models.  A unique solution can then be identified by constraining the minimal element in each row of $\mathbf{P}^{j}$ (and thus $\mathbf{Z}^{j}$) to be zero for all $j$. See Fig.~\ref{fig:MTDmLTDidentifiability} for an illustration. The intuition for this result is simple: any excess probability mass on a row of each $\mathbf{Z}^{j}$ may be pushed onto the same row of the intercept term $\mathbf{z}^{0}$ without changing the full conditional probability. 

The above identifiability condition also provides interpretation for the parameters in the MTD model. Specifically, the element $\mathbf{Z}_{m n}^{j}$ denotes the additive increase in probability that $x_{ti}$ is in state $m$ given that $x_{(t-1)j}$ is in state $n$. Furthermore, the  $\gamma_{j}$ parameters now represent the total amount of probability mass in the full conditional distribution explained by categorical variable $x_j$, providing an interpretable notion of dependence in categorical time series. 

Unfortunately, the set of $\mathbf{Z}^{j}$ that satisfy the MTD identifiability constraints is non-convex since the locations of the zeros are unknown.  \citet{tank2021categorical} address this issue by adding a penalty $\Omega(\mathbf{Z})$ that biases the solution towards the uniqueness constraints. This regularization also aids convergence of optimization since the maximum likelihood solution without identifiability constraints is not unique. The regularized estimation problem is given by
\begin{align}
\label{eq:3.5}
\begin{array}{l}{\underset{\mathbf{Z}, \gamma}{\text { minimize }} L_{\mathrm{MTD}}(\mathbf{Z})+\lambda \Omega(\mathbf{Z})} \\ {\text { subject to } \mathbf{1}^{T} \mathbf{Z}^{j}=\gamma_{j} \mathbf{1}^{T},\quad \mathbf{Z}^{j} \geq 0, \forall j, \quad \mathbf{1}^{T} \gamma=1, \gamma \geq 0.}\end{array}
\end{align}

As \citet{tank2021categorical} show, for any $\lambda>0$ and $\Omega(\mathbf{Z})$ not dependent on $\mathbf{z}^{0}$ and increasing with respect to the absolute value of entries in $\mathbf{Z}^{j}$, the solution to the problem in (\ref{eq:3.5}) is contained in the set of identifiable MTD models. 
Intuitively, by penalizing the entries of the $\mathbf{Z}^{j}$ matrices, but not the intercept term, solutions will be biased to having the intercept contain the excess probability mass, rather than the $\mathbf{Z}^{j}$ matrices. An entire class of regularizers match the necessary conditions and can be considered.

\begin{proposition}\label{prop:identifiableMTDGranger}\citep{tank2021categorical}
Based on the MTD identifiability constraint where each row must have at least one zero element, $x_j$ is Granger non-causal for $x_i$ if and only if $\mathbf{Z}^{j}=0$ (a special case of all columns being equal).
\end{proposition} 

To both enforce the identifiability constraints and select for Granger non-causality, \citet{tank2021categorical} explore a set of penalties $\Omega(\mathbf{Z})$ that encourage some $\mathbf{Z}^{j}$ to be zero, while maintaining convexity of the overall objective.  These penalties include an $L_1$ penalty on the $\gamma_j$ (with $\gamma_j$=0 implying $\mathbf{Z}^j=0$); a group lasso penalty on each $\mathbf{Z}^j$~\citep{yuan2006}; and a group-lasso-type penalty that scales with the number of categories per series, $m_j$, to avoid differentially penalizing series based on their number of categories.
To solve the penalized estimation problem, \citet{tank2021categorical} developed both projected gradient and Frank-Wolfe algorithms for the MTD model that harnesses the convex formulation. For the projected gradient optimization, they further developed a Dykstra projection method to quickly project onto the MTD constraint set, allowing the MTD model to scale to
much higher dimensions.

\subsubsection{Alternative formulation for categorical time series}
\citet{tank2021categorical} also propose a multinomial logistic transition distribution model (mLTD) as an alternative to the MTD:
\begin{equation}
\label{eq:2.6}
p\left(x_{ti} | x_{(t-1)1}, \ldots, x_{(t-1)p}\right)=\frac{\exp \left(\mathbf{z}_{x_{ti}}^{0}+\sum_{j=1}^{p} \mathbf{Z}_{x_{ti}, x_{(t-1)j}}^{j}\right)}{\sum_{x^{\prime} \in \mathcal{X}_{i}} \exp \left(\mathbf{z}_{x^{\prime}}^{0}+\sum_{j=1}^{p} \mathbf{Z}_{x^{\prime}, x_{(t-1)j}}^{j}\right)},
\end{equation}
where $\mathbf{Z}^{j} \in \mathbb{R}^{m_{i} \times m_{j}}$ and $\mathbf{z}^{0} \in \mathbb{R}^{m_{i}}$. As with the MTD, interaction terms may be added. Granger causality follows identically to the MTD case in Proposition~\ref{Proposition 3.3}: \emph{$x_j$ is Granger non-causal for $x_i$ iff the columns of $\mathbf{Z}^j$ are all equal}. 

The non-identifiability of multinomial logistic models is well-known, as is the non-identifiability of generalized linear models with categorical covariates. Combining the standard identifiability restrictions for both settings clarifies that every mLTD has a unique parameterization such that first column and last row of $\mathbf{Z}^{j}$ are zero for all $j$ and the last element of $\mathbf{z}^{0}$ is zero \citep{Agresti2011Categorical}.
See Fig.~\ref{fig:MTDmLTDidentifiability}.  Although the mLTD identifiability conditions differ from those of the MTD, Granger non-causality interpretation of the identifiable mLTD mirrors the identifiable MTD in Proposition~\ref{prop:identifiableMTDGranger}: \emph{$x_j$ is Granger non-causal for $x_i$ iff $\mathbf{Z}^{j}=0$ (a special case of all columns being equal).}

To select for Granger causality in the mLTD model while enforcing identifiability, akin to the MTD case, \cite{tank2021categorical} propose a group lasso penalty on each of the $\mathbf{Z}^{j}$ matrices, leading to the following optimization problem:
\begin{equation}
\begin{aligned}
&\underset{\mathbf{Z}}{\operatorname{minimize}} \sum_{t=1}^{T} \mathbf{z}_{x_{ti}}^{0}+\sum_{j=1}^{d} \mathbf{Z}_{x_{ti} x_{(t-1)j}}^{j} \\
&\quad \quad+\log \left(\sum_{x^{\prime} \in \mathcal{X}_{i}} \exp \left(\mathbf{z}_{x^{\prime}}^{0}+\sum_{j=1}^{d} \mathbf{Z}_{x^{\prime} x_{(t-1)j}}^{j}\right)\right)+\lambda \sum_{j=1}^{d}\left\|\mathbf{Z}^{j}\right\|_{F} \\
&\text { subject to } \quad \mathbf{Z}_{1 : m_{i}, 1}^{j}=0, \mathbf{Z}_{m_{i}, 1 : m_{j}}^{j}=0\ \forall j.
\end{aligned}
\end{equation}
For two categories, $m_{i}=2\ \forall i$, this problem reduces to sparse logistic regression for binary time series, which was studied by \citet{hall2016inference}. As in the MTD case, the group lasso penalty shrinks some $\mathbf{Z}^{j}$ entirely to zero.

Although the MTD and mLTD are conceptually similar, the parameters of the mLTD are unfortunately harder to interpret.  Another alternative formulation one might consider is based on the MTD-probit model of \citet{nicolau2014new}; however, this framework is not a natural fit for inferring Granger causality,  both due to the non-convexity of the probit model and the non-convex constraints on $\mathbf{Z}^{j}$ matrices. 

\subsubsection{Estimating networks of binary and count time series}
The MTD and mLTD models are specifically geared for Granger causal analysis of autoregressive categorical processes.  \cite{hall2016inference} instead study a broad class of generalized linear autoregressive (GLAR) models, capturing Bernoulli and log-linear Poisson autoregressive (PAR) models, and focus on the high-dimensional multivariate setting.  The GLAR model is specified as:
\begin{align}
    x_{ti} \mid \x_{<t} \sim p(\nu_i + \mathbf{a}_i^\top \x_{<t}),
    \label{eq:GLAR}
\end{align}
where $p$ is an exponential family probability distribution. 
The formulation in (\ref{eq:GLAR}) follows a component-wise structure, and from Definition~\ref{def:strongGranger} we can decipher that \emph{time series $x_j$ does not Granger cause series $x_i$ iff $a_{ij}=0$}. 

\citet{hall2016inference} consider $L_1$ regularization of $A$ constructed row-wise from $\mathbf{a}_i$.  They derive statistical guarantees, such as sample complexity bounds and mean-squared error bounds 
for the sparsity-regularized maximum likelihood estimator, addressing the key challenge of correlations and potential heteroscedasticity in the GLAR observations.

Count data can also be analyzed using autoregressive models with thinning operators of previous counts---so-called integer-valued autoregressive (INAR) processes \citep{mckenzie2003discrete,Weiss2018}.  One example is the Poisson INAR, which performs binomial thinning and adds Poisson innovations. In the univariate case, the process has Poisson margins; in the multivariate case, although a stationary distribution exists, the margins are no longer Poisson unless the thinning matrix is diagonal. \citet{AldorNoiman2016} capture dependence between the dimensions of a multivariate count process through the Poisson rate parameters of a multivariate Poisson INAR with diagonal thinning, using multiple shrinkage via a Dirichlet process prior on the rate parameters.  The resulting clustering of count time series gives a (strict) notion of Granger non-causality for any pair of series appearing in disjoint clusters.  

Another approach is the INGARCH model~\citep{Weiss2018}, which leverages an AR-like model on the conditional mean $\mathbf{M}_t = \mathbb{E}[\x_t\mid \x_{<t}] = \alpha_1 \x_{t-1} + \beta_0$, and is useful for modeling overdispersed counts. One example is modeling Poisson-distributed counts with rate parameter defined via the conditional mean process $\mathbf{M}_t$; other specifications consider binomial or negative binomial conditional distributions.  The INGARCH model has connections both to the GLAR of (\ref{eq:GLAR}), as well as to the popular GARCH model \citep[see, e.g,][]{BauwensLaurentRombouts2006}.  However, the INGARCH model has most commonly been used in low-dimensional settings, often univariate, and scaling the model to higher dimensional settings and using it for Granger causality analysis is an open research area, as with the Poisson INAR.

\subsubsection{Granger causal interactions in point processes}
A key assumption of the standard Granger causal framework is that observations are on a fixed, discrete-time grid. In Section~\ref{sec:mixedFreqGranger}, we consider cases where the sampling rate 
might not match the time scale of the true causal interactions.  Here we focus on another important case emerging from irregularly and asynchronously observed time series better modeled via point processes in continuous time.  

Inferring Granger causal interactions in the general class of multivariate point processes is often challenging due to intractability of representing the histories of the processes and their impact on the process' evolution.  Recent work gained traction by focusing specifically on Hawkes processes, describing self- and mutually-excitatory processes~\citep{zhou2013learning,eichler2017graphical,xu2016learning}.  Early applications of Hawkes processes include modeling seismic activity and neural firing patterns, with more recent applications to interactions in social networks and medical event streams.  For Granger causality analysis, \citet{eichler2017graphical} provide straightforward conditions on the link functions of the conditional intensities of the multivariate Hawkes process and derive a nonparametric estimation procedure. 

Let $N = \{N(t)|t \in [0, T]\}$ be a point process arising from a Hawkes process with conditional intensity functions
\begin{align}
    \lambda_i(t) = \nu_i + \sum_{j=1}^p \phi_{ij}(u)dN_j(t-u) \quad i=1,\dots, p,
\end{align}
where $\nu_i$ is the baseline intensity and $\phi_{ij}$ are the link functions with $\phi_{ij}(u)=0$ for $u\leq 0$ and $\int_0^\infty \|\phi_{ij}(u)\|du \leq 1$.  Then, \emph{$N_j$ does not Granger cause $N_i$ if and only if $\phi_{ij}(u) = 0$ for all $u \in \mathbb{R}$} \citep{eichler2017graphical}. 

\citet{zhou2013learning}, \citet{xu2016learning} and \citet{hansen2015lasso} recently used sparsity-inducing penalties to infer (high-dimensional) Granger causal networks from Hawkes processes. Motivated by neuroscience applications, \citet{chen2017hawkes2} generalized Hawkes processes to allow for inhibitory interactions, \citet{chen2017hawkes1} proposed a screening approach for efficient estimation of high-dimensional Hawkes process networks, and \citet{wang2020statistical} developed a high-dimensional inference framework for Hawkes processes. 
The PAR model version of (\ref{eq:GLAR}) is also closely related to the continuous-time Hawkes process model, and can be used as an alternative to the above approaches.

\subsection{Methods for capturing interactions in non-linear time series}
\label{sec:nonlinear}

Beyond the analysis of discrete-valued time series, as in Section~\ref{sec:categorical}, there are a range of other scenarios where the relationships between the past of one series and future of another falls outside of the VAR model class of traditional model-based Granger causailty analysis.  In such cases, model-based methods have been shown to fail in numerous real-world settings~\citep{terasvirta2010modelling, tong2011nonlinear, lusch2016inferring}.  
One example is time series with heavy tails, which has been modeled using VARs with elliptical errors \citep{qiu2015robust}. Another example of particular importance in a number of applications---and one we focus on in this review---is that of nonlinear interactions. Model-free methods, like transfer entropy \citep{vicente2011transfer} or directed information \citep{amblard2011directed}, can detect nonlinear dependencies between past and future with minimal assumptions on the predictive relationships. However, these estimators have high variance and require large amounts of data for reliable estimation. These approaches also suffer from curse of dimensionality \citep{runge2012escaping}, making them inappropriate in high-dimensional settings.

Dynamical system representations, often in the form of coupled ordinary differential equations (ODEs),  have long been used to capture nonlinear relationship in time series. While ODEs are inherently deterministic, a commonly-used approach is to assume that data from the underlying ODEs are contaminated with mean-zero additive noise $\e_t$:
\begin{align}
    \dot x_{ti} &= \alpha_i + f_i(\x_t) \label{eqn:generalode1} \\ 
    \y_t &= \x_t + \e_t, \label{eqn:generalode2}
\end{align}
where $f_i: \mathbb{R}^p \to \mathbb{R}$ is a function mapping the current state of all variables to the change in $x_i$ (the derivative $\dot x_{ti}$). 

While ODE-based approaches for analyzing specific systems use parametric forms, more recent work has focused on system identification using flexible specifications of functions $f_i$. One such approach, which has been successfully applied to high-dimensional problems is to consider an \emph{additive} ODE instead of (\ref{eqn:generalode1}); that is,
\begin{align}\label{eqn:additiveode}
    \dot x_{ti} &= \alpha_i + \sum_{j=1}^p f_{ij}(x_{tj}).
\end{align}
For the system in (\ref{eqn:additiveode}), it follows from Definition~\ref{def:granger_g} that \emph{$x_j$ is Granger non-causal for $x_i$ if and only if $f_{ij} = 0$}. Using this connection, \citet{henderson2014} and \citet{wu2014} developed regularized nonparametric estimation procedures to infer nonzero functions, $f_{ij}$, and  \citet{chen2017ode} addressed the key challenge of estimating the derivative $\dot x_{ti}$ and established the consistency of the network Granger causality estimates.

The ODE-based approaches discussed above offer flexible alternatives to parametric approaches for modeling nonlinear dynamics. However, they are limited to additive interaction mechanisms. 
A promising alternative is to consider more general dynamics and interactions by leveraging neural networks. Neural networks can represent complex, nonlinear, and non-additive interactions between inputs and outputs. Indeed, their time series variants, such as autoregressive multilayer perceptrons (MLPs) \citep{raissi2018multistep, kicsi2004river, billings2013nonlinear} and recurrent neural networks (RNNs) like long-short term memory networks (LSTMs) \citep{graves2012supervised} have shown impressive performance in forecasting multivariate time series given their past \citep{yu2017long, zhang2003time, li2017graph}.

Consider a \textit{nonlinear} autoregressive (NAR) model that allows ${\bf x}_t$ to evolve according to general nonlinear dynamics \citep{billings2013nonlinear}, assuming an additive zero mean noise $\e_t$
\begin{align}
{\bf x_t} =g(x_{<t1} ,\ldots,x_{<tp} )+ \e_t.
\end{align}
In a NAR \emph{forecasting} setting, there is a long history of modeling $g$ using neural networks, via both traditional architectures \citep{billings2013nonlinear,chu1990neural,billings1996determination} and more recent deep learning techniques \citep{yu2017long,li2017graph,tao2018hierarchical}. These approaches either utilize an MLP with inputs $\x_{<t} = \x_{(t-1):(t-K)}$, for some lag $K$, or a recurrent network, like an LSTM, that does not require specifying the lag order.

While these methods have shown impressive predictive performance, they are essentially black box models and provide little interpretation of the multivariate structural relationships in the series. In the context of Granger causality, due to sharing of hidden layers, it is difficult to specify sufficient conditions on the weights that simultaneously allows series $j$ to Granger cause series $i$ but not other series $i'$ for $i \neq i'$. A second drawback is that jointly modeling a large number of series leads to many network parameters. Thus, these methods require much more data to fit reliably and tend to perform poorly in high-dimensional settings. Finally, a joint network over all $x_{ti}$ for all $i$ assumes that each time series depends on the same past lags of other series. However, in practice, each $x_{ti}$ may depend on different past lags of other series. As in linear methods discussed in Section~\ref{sec:penalizedngc}, appropriate lag selection is crucial for Granger causality selection in nonlinear approaches---especially in highly parametrized models like neural networks. 

With an eye towards inferring Granger causality, but simultaneously tackling the sample complexity and lag selection problems, \citet{tank2018neural} propose a framework leveraging the component-wise model of (\ref{eq:compVAR}) that disentangles the effects of lagged inputs on individual output series. The method models the component-wise transition functions $g_i$ using neural networks---either via an MLP or RNN like the LSTM---and deploys carefully constructed sparsity-inducing penalties on particular groupings of neural network weights to identify Granger non-causal interactions.  One of the penalties---building on the hierarchical group lasso \citep{nicholson2017varx,huang2011learning,kim2010tree}---automatically detects both nonlinear Granger causality and also the lags of each inferred interaction in the MLP setting.  The LSTM-based formulation, on the other hand, sidesteps the lag selection problem entirely because the recurrent architecture efficiently models long-range dependencies \citep{graves2012supervised}. The proposed penalties also aid in handling limited data in the high-dimensional setting.  We review each approach below.

\begin{figure}[t]
\centering
\includegraphics[width=.8\linewidth, clip=TRUE, trim=0cm 0cm 0cm 0cm]{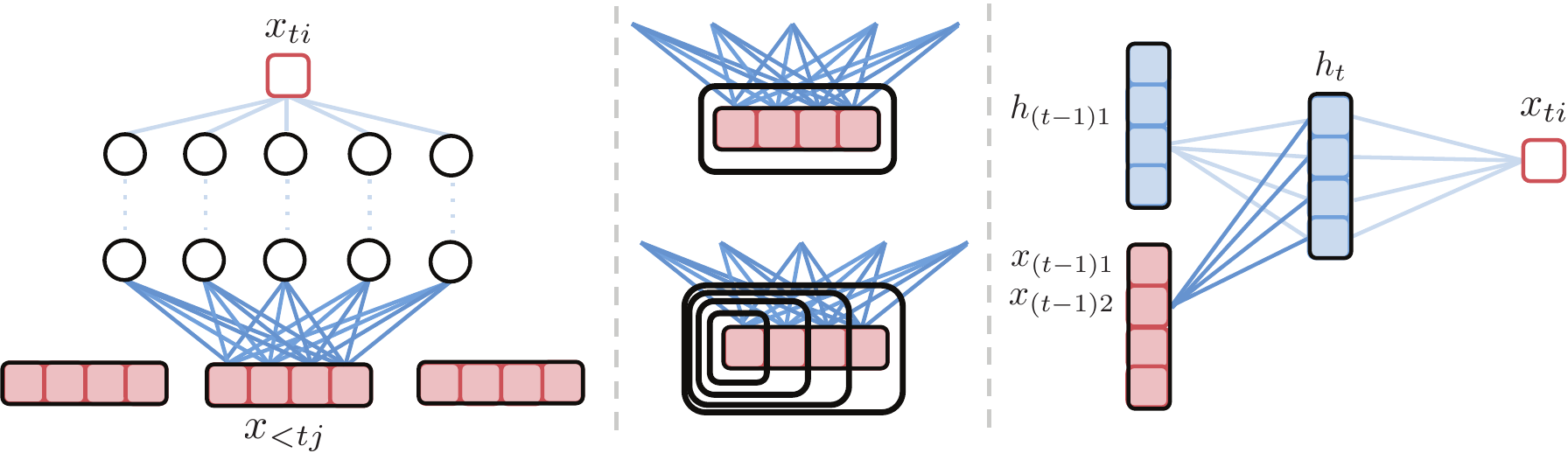}
\caption{(\emph{left}) Schematic for cMLPs. If outgoing weights for $x_{<tj}$ (dark blue) are penalized to zero, then $x_j$ does not Granger-cause $x_i$. (\emph{center}) The group lasso penalty jointly penalizes the full set of outgoing weights while the hierarchical version penalizes the nested set of outgoing weights, penalizing higher lags more. (\emph{right}) Schematic for cLSTM. If outgoing weights to hidden units from an input $x_{(t-1) j}$ are zero, then $x_j$ does not Granger-cause $x_i$.}
\label{Fig1}
\end{figure}

\subsubsection{MLP}
\begin{figure}[t]
\centering
\includegraphics[page=8,width=0.95\textwidth, clip=TRUE, trim=0cm 10.5cm 0cm 3cm]{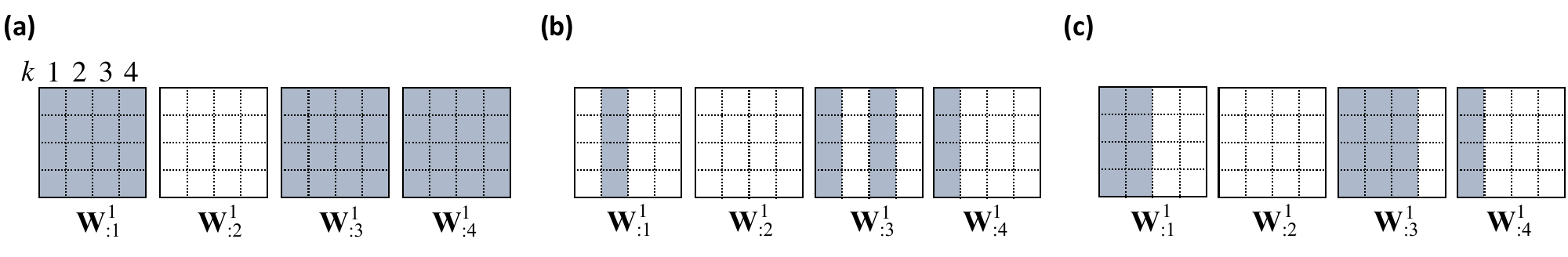}    
\caption{Example of group sparsity patterns of cMLP first layer weights with four first layer hidden units ($H=4$) and four input series ($p=4$) with maximum
lag $k = 4$. Differing sparsity patterns are shown for the three different
structured penalties: (\emph{a}) group lasso (\texttt{GROUP}), (\emph{b}) group
sparse group lasso (\texttt{MIXED}) and (\emph{c}) hierarchical lasso
(\texttt{HIER}).}
\label{Fig2}
\end{figure}

Define $g_i$ via an MLP with $L-1$ layers and $\h_{t}^{l}$ representing the $H$ values of $l$th hidden layer at time $t$. The parameters are given by weights $W^l$ and biases $\mathbf{b}^l$ at each layer (with appropriate dimensions for that layer). To draw an analogy with the linear VAR model of (\ref{eqn:var}), we further decompose the weights at the first layer across time lags, $W^{1}=\left\{W^{11}, \ldots, W^{1 K}\right\}$.  The resulting \textit{component-wise} MLP (cMLP) is given as \citep{tank2018neural}:
\begin{equation}
    \label{nn_layer}
\begin{aligned}  
\h_{t}^{1}&=\sigma\left(\sum_{k=1}^{K} W^{1 k} \mathbf{x}_{t-k}+\mathbf{b}^{1}\right)\\
\h_{t}^{l}&=\sigma\left(W^{l} \h_{t}^{l-1}+\mathbf{b}^{l}\right) \quad l=2,\dots,L-1\\
x_{t i}&=W^{L} \h_{t}^{L-1}+b^{L}+e_{t i},
\end{aligned}
\end{equation}
where $\sigma$ is an activation function, such as \texttt{logistic} or \texttt{tanh}, and $e_{ti}$ is mean zero white noise. In \cite{tank2018neural}, a linear output decoder $W^L$ is used. 
However, as the authors mention, other decoders like a \texttt{logistic}, \texttt{softmax}, or Poisson likelihood with exponential link function \citep{mccullagh1989generalized}, could be used to model nonlinear Granger causality in multivariate binary \citep{hall2016inference}, categorical \citep{tank2021categorical}, or positive count time series \citep{hall2016inference}. From (\ref{nn_layer}), the Granger non-causality conditions are straightforward to elicit:
\begin{proposition}\label{prop:GrangerMLP} \citep{tank2018neural}
In the MLP model of (\ref{nn_layer}), following Definition~\ref{def:granger_g}, if the $j$th column of the first layer weight matrix, $W_{ : j}^{1 k}$, contains zeros for all $k$, then series $x_j$ does not Granger-cause series $x_i$. 
\end{proposition}
By Proposition~\ref{prop:GrangerMLP}, if the first layer weight matrix, $W_{ : j}^{1 k}$, contains zeros for all $k$ then $x_{<tj}$ does not influence the hidden unit $h_{t}^{1}$ and thus the output $x_{t i}$. Following Definition \ref{def:granger_g}, we see that $g_i$---which is implicitly defined through the hidden layers of the MLP in (\ref{nn_layer})---is then invariant to $x_{<tj}$. Thus, analogously to the VAR case, one may select for Granger causality by applying a group penalty to the columns of the $W^{1 k}$ matrices for each $g_i$,
\begin{equation} \label{nn_pen}
\min _{\mathbf{W}} \sum_{t=K}^{T} \Big(x_{i t}-g_{i}\left(\x_{(t-1) :(t-K)}\right) \Big)^2+\lambda \sum_{j=1}^{p} \Omega\left(W_{ : j}^{1}\right),
\end{equation}
where $\Omega$ is a penalty that shrinks the entire set of first layer weights for input series $j$, i.e., $W_{ : j}^{1}=\left(W_{ : j}^{11}, \ldots, W_{ : j}^{1 K}\right)$, to zero. Three penalties, illustrated in Fig.~\ref{Fig2}, are considered by \citet{tank2018neural}: (i) a group lasso penalty over the entire set of outgoing weights across all lags for time series $j$, $W_{ : j}^{1}$ (the analogue to the group lasso penalty across lags in the VAR case); (ii) a novel \textit{group sparse group lasso} penalty that provides both sparsity across groups (a sparse set of Granger causal time series) and sparsity within groups (a subset of relevant lags); and (iii) a \textit{hierarchical} group lasso penalty to simultaneously select for both Granger causality and the lag order of the interaction. 

\subsubsection{RNN} 
As in the MLP case, it is difficult to disentangle how each series affects the evolution of another series when using a standard RNN. This problem is even more severe in complicated recurrent networks like LSTMs.  For a general RNN, the hidden state at time $t$ is updated recursively:
\begin{equation}
\label{eq:hidden}
\begin{aligned} 
{\bf h}_t &= f_i({\bf x}_t, {\bf h}_{t-1})\\
x_{ti} &= W^2 {\bf h}_t + e_{ti},
\end{aligned}
\end{equation}
where $f_i$ is a nonlinear function that depends on the particular recurrent architecture and $W^2$ are the output weights.  

Due to their effectiveness at modeling complex time dependencies, \citet{tank2018neural} focus on modeling the recurrent function $f_i$ using an LSTM \citep{graves2012supervised}. The LSTM introduces a second hidden state variable ${\bf c}_t$, the \emph{cell state}, and updates its set of hidden states $({\bf c}_t , {\bf h}_t )$ recursively as
\begin{equation}
\label{eq:lstm}
\begin{aligned} 
{\bf f}_t &= \sigma \left(W^f {\bf x}_t + U^f {\bf h}_{(t - 1)} \right) \\
{\bf i}_t &= \sigma \left(W^{in} {\bf x}_t + U^{in} {\bf h}_{(t - 1)} \right) \\
{\bf o}_t &= \sigma \left(W^{o} {\bf x}_t + U^{o} {\bf h}_{(t - 1)} \right) \\
{\bf c}_t &= {\bf f}_t \odot {\bf c}_{t-1} + {\bf i}_t \odot \sigma \left(W^c {\bf x}_t + U^c {\bf h}_{t - 1} \right) \\
{\bf h}_t &= {\bf o}_t \odot \sigma ({\bf c}_t),
\end{aligned}
\end{equation}
where $\odot$ denotes element-wise multiplication.  The input (${\bf i}_t$), forget (${\bf f}_t$), and output (${\bf o}_t$) gates control how each component of the cell state (${\bf c}_t$) is updated and then transferred to the hidden state (${\bf h}_t$) used for prediction.  The additive form of the cell state update in the LSTM allows it to encode long-range dependencies: cell states from far in the past may still influence the cell state at time $t$ if the forget gates remain close to one. In the context of Granger causality, this flexible architecture can represent long-range, nonlinear dependencies between time series. 

\begin{figure}[t]
\centering
\includegraphics[page=10,width=0.5\textwidth, clip=TRUE, trim=5cm 10cm 12cm 3cm]{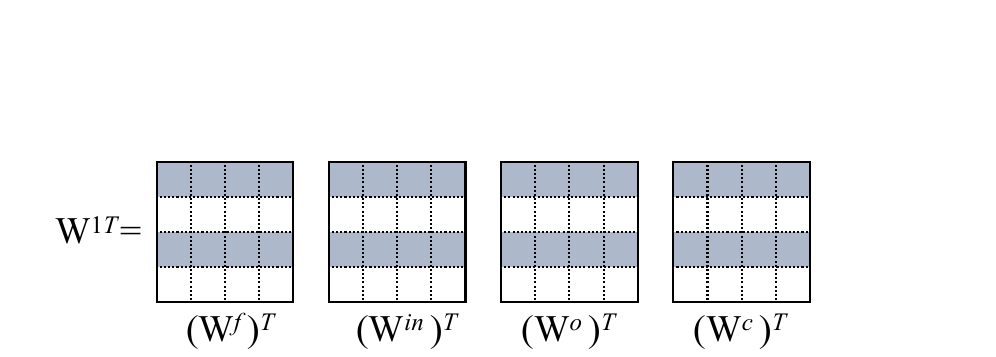}    
\caption{Example of group sparsity patterns in a cLSTM with $H=4$ and $p=4$. 
Due to the group lasso penalty on the columns of $W$, the $W^f$ , $W^{in}$, $W^o$, and $W^c$ matrices share the same column sparsity pattern.}
\label{Fig3}
\end{figure}

Let ${\bf W} = (W^1, W^2, U^1)$ be the full set of parameters, where $W^1 = \left( (W^f)^\top,(W^{in})^\top ,(W^o)^\top ,(W^c)^\top \right) ^\top$ and $U^1 = \left( (U^f)^\top,(U^{in})^\top ,(U^o)^\top ,(U^c)^\top  \right)^\top$ are the full set of first layer weights. In (\ref{eq:lstm}), the set of input matrices $W^1$ controls how the past time series affect the hidden representation update, and thus the prediction of $x_{ti}$. Granger non-causality for this \emph{component-wise LSTM model} (cLSTM) follows directly from Definition~\ref{def:granger_g}:

\begin{proposition}\label{prop:GrangerLSTM} \citep{tank2018neural}
For the cLSTM of (\ref{eq:hidden})--(\ref{eq:lstm}), following Definition~\ref{def:granger_g}, a sufficient condition for Granger non-causality of a series $x_j$ on a series $x_i$ is that all elements of the $j$th column of $W^1$ are zero, $W^1_{:j} = 0$.
\end{proposition}

Thus, we may select for Granger causality using a group lasso penalty across columns of $W^1$ and considering: 
\begin{align} \label{eq:nn_group_lasso}
\min_{\mathbf{W}} \sum_{t = 2}^\top \Big(x_{it} - g_{i}(\x_{<t}) \Big)^2 + \lambda \sum_{j = 1}^p \|W^1_{:j}\|_2.
\end{align}
As with the cMLP, $g_i$ for the cLSTM is implicitly defined through the recurrent structure of (\ref{eq:hidden})--(\ref{eq:lstm}). For larger $\lambda$s, many columns of $W^1$ will be zero, leading to a sparse set of Granger causal connections; see Fig.~\ref{Fig3}. 
\citet{tank2018neural} optimize the objectives in (\ref{nn_pen}) and (\ref{eq:nn_group_lasso}) (under various choices of penalty) using proximal gradient descent.

\subsection{Subsampled and mixed-frequency time series}
\label{sec:mixedFreqGranger}

\begin{figure}
\centering
\captionsetup{width=\linewidth}
\includegraphics[page=11,width=0.9\textwidth, clip=TRUE, trim=0cm 8.2cm 0cm 6.2cm]{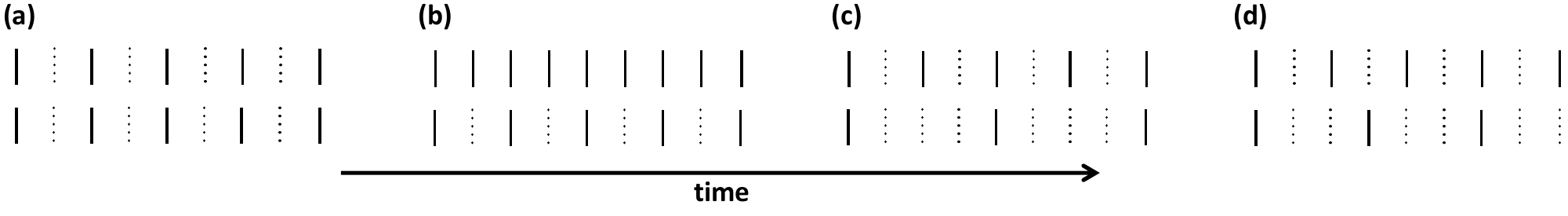}    
\caption{Four types of structured sampling. Black lines indicate observed data and dotted lines indicate missing data. (a) Both series are subsampled. (b) The standard mixed-frequency case, where only the second series is subsampled. (c) A subsampled version of (b) where each series is subsampled at different rates. (d) A subsampled mixed-frequency series that has no common factor across sampling rates and thus is not a subsampled version of (b). } \label{sampling_types}
\end{figure}

Even if the time series follows a linear VAR (\ref{eqn:var}), if the process is observed at a sampling rate slower than the true causal scale of the underlying process, as depicted in Fig.~\ref{sampling_types}(a), a causal analysis rooted at this slower time scale may miss true interactions and add spurious ones \citep{zhou:2014,silvestrini:2008,boot:1967,breitung:2002}. Mixed frequency time series also present a challenge to Granger causal analysis.  Example scenarios are depicted in Fig.~\ref{sampling_types}(b)-(d). Scenario (b) often arises in econometrics, amongst other fields, and VAR models are fit at the scale of the least finely sampled time series~\citep[see e.g.,][]{schorfheide:2015}. However, for macroeconomic indicators like GDP, the scale of sampling is generally arbitrary and may not reflect the true causal dynamics, leading to confounded Granger and instantaneous causality judgments \citep{zhou:2014,breitung:2002}. Scenarios (c)-(d) combine subsampled and mixed frequency settings and their respective challenges.

Recently, causal discovery in subsampled time series has been studied with methods in causal structure learning using graphical models \citep{danks:2013,plis:2015,hyttinen:2016}. These methods are model-free and automatically infer a sampling rate for causal relations most consistent with the data.  For mixed-frequency autoregressive models with no subsampling at the fastest scale, Fig.~\ref{sampling_types}(b), finding identifiability conditions was an open problem for many years \citep{chen:1998}. \citet{anderson:2015} recently showed that in Scenario (b), a \emph{non-structural} autoregressive model is generically identifiable from the first two observed moments, so unidentifiable models make up a set of measure zero of the parameter space.  In this section, we instead outline the model-based approach and identifiability conditions explored by~\citet{tank2019biometrika} for Granger causal analysis of structural vector autoregressive (SVAR) models under both subsampling and mixed-frequency settings.

An SVAR~\citep{lutkepohl2005} allows the dynamics of $\x_t$ to follow a combination of instantaneous effects, autoregressive effects, and independent noise.  For simplicity, let us consider a lag one SVAR:
\begin{align} \label{eq:SVAR}
\x_t &= \B \x_t + \D \x_{t-1} + \e_t,
\end{align}
where $\B \in \mathbb{R}^{p \times p}$ is the structural matrix that determines the instantaneous linear effects, $\D \in \mathbb{R}^{p \times p}$ is an autoregressive matrix that specifies the lag one effects conditional on the instantaneous effects, and $\e_t \in \mathbb{R}^{p}$ is a white noise process such that $E(\e_t) = 0 \,\,$ for all $t$, and $e_{ti}$ is independent of $e_{t'j}$ for all $i, j,t,t'$ such that $(i,t) \neq (j,t)$. We assume $e_{tj}$ is distributed as  $e_{tj} \sim p_{e_j}$. Solving (\ref{eq:SVAR}) in terms of $\x_t$ gives  the following lag one SVAR process: 
\begin{align}
\x_t &= (I - \B)^{-1} \D \x_{t-1} + (I - \B)^{-1} \e_t = \A \x_{t-1} + \C \e_t. \label{eq:svar_f}
\end{align}
In (\ref{eq:svar_f}), $\A_{ij}$ denotes the lag one linear effect of series $x_j$ on series $x_i$ and $\C \in \mathbb{R}^{p \times p}$ is the structural matrix. The error $e_{ti}$ is known as the shock to series $x_i$ at time $t$ and the element $\C_{ij}$ is the linear instantaneous effect of $e_{tj}$ on $x_{ti}$.  The most typical condition is that $\C$ is lower triangular with ones on the diagonal, implying a known causal ordering of the instantaneous effects. 
When the errors, $\e_t$, are non-Gaussian, both the causal ordering and instantaneous effects $\C$ may be inferred directly from the data using techniques from independent component analysis \citep{hyvarinen:2010}. Alternatively, $\C$ can be directly estimated via maximum likelihood \citep{lanne:2015}.

In the subsampled case, Fig.~\ref{sampling_types}(a), we observe $\x_t$ every $k$ time steps leading to  
${\bf \tilde{X}} = \left(\tilde{\x}_1, \tilde{\x}_2, \ldots, \tilde{\x}_{\tilde{T}}\right) \equiv \left(\x_1, \x_{1 + k},\ldots,\x_{1 + (\tilde{T}-1)k}\right)$ 
observations, where $\tilde{T}$ is the number of subsampled observations. By marginalizing out the unobserved $\x_t$, we obtain the evolution equations
\begin{align}
\tilde{\x}_{t} &= \x_{1 + tk} = \A \x_{1 + tk - 1} + \C \e_{1 + tk} = \A \left( \A \x_{1 + tk - 2} + \C \e_{1 + tk - 1} \right) + \C \e_{1 + tk} \nonumber \\
&= (\A)^k \tilde{\x}_{t-1} + \sum_{l = 0}^{k-1} (\A)^l \C \e_{1 + tk - l} \label{eq:svar_sub_f}\\
&= (\A)^k \tilde{\x}_{t-1} + \Ll \tilde{\e}_t \label{eq:svar_sub},
\end{align} 
where $\tilde{\e}_t = \left(\e_{1 + tk}^\top, \ldots \e_{2 + (t - 1)k}^\top\right)^\top$ is the stacked vector of errors for time $1 + tk$ and the unobserved points between $1 + tk$ and $1 + (t - 1)k$ and $\Ll = \left(\C, \ldots, (\A)^{k-1} \C\right)$. Eq.~(\ref{eq:svar_sub_f}) states that the subsampled process is a linear transformation of the past subsampled observations with transition matrix $(\A)^{k}$ and a weighted sum of the shocks across all unobserved time points. Each shock is weighted by $\A$ raised to the power of the time lag. 
Eq.~(\ref{eq:svar_sub}) appears to take a similar form to the structural process in (\ref{eq:SVAR}); however, now the vector of shocks, $\tilde{\e}_t$, is of dimension $kp$, with special structure on both the structural matrix $\Ll$ and the distributions of the elements in $\tilde{\e}_t$. Unfortunately, this representation does not have the interpretation of instantaneous causal effects, as there are now multiple shocks per individual time series. We will refer to the full parametrization of the subsampled structural model in (\ref{eq:svar_sub}) as $\left(\A, \C, p_e; k\right)$. 

\begin{figure} 
\centering 
\captionsetup{width=\linewidth}
\includegraphics[width=.65\textwidth]{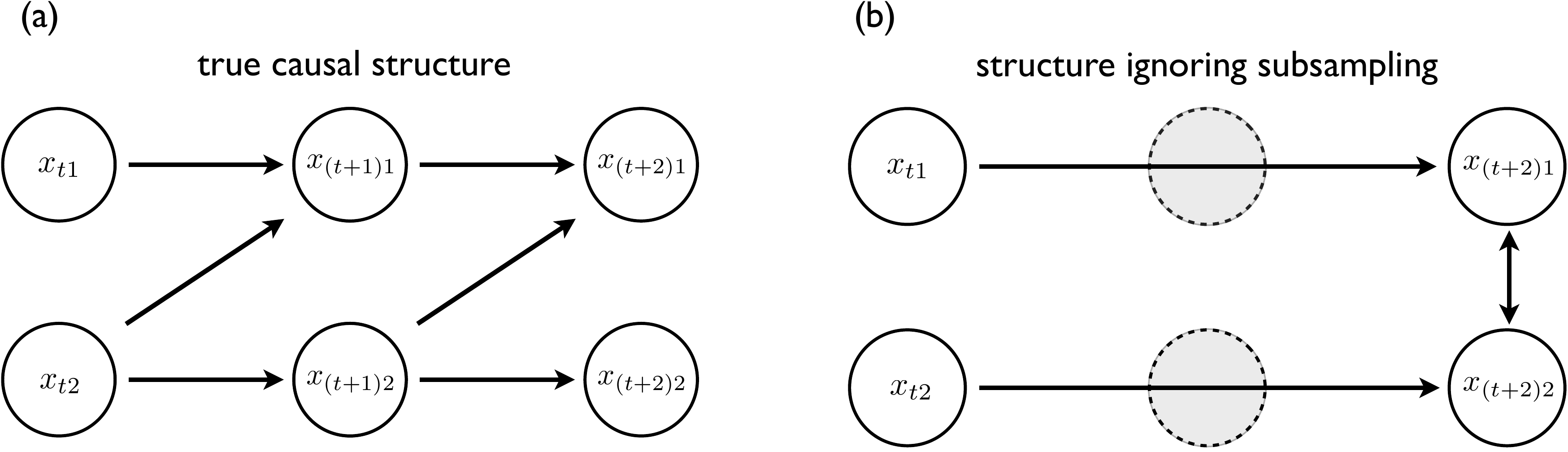}
\caption{
Depiction of how subsampling confounds causal analysis of lagged and instantaneous effects. (a) True causal diagram for regularly sampled data. (b) Estimated causal structure when subsampling is ignored.} \label{confounds}
\end{figure}

A classical analysis based on $\tilde{\x}_t$ that does not account for subsampling would incorrectly estimate lagged Granger causal effects in $(\A)^k$, because $\A_{ij} = 0$ does not imply that $((\A)^k)_{ij} = 0$, and vice versa \citep{Gong:2015}. Similarly, estimation of structural interactions may also be biased if subsampling is ignored. This is illustrated in Fig.~\ref{confounds}, where an analysis based on subsampled data 
identifies no lagged causal effect between $x_1$ and $x_2$, but a relatively large instantaneous interaction. See \citet{tank2019biometrika} for further details and examples.

The mixed frequency scenarios, Fig.~\ref{sampling_types}(b)-(d), are also considered by \citet{tank2019biometrika} and involve defining sampling rates for each series and a set of indicator matrices that select the observed time points from (\ref{eq:svar_f}).  Despite more cumbersome notation, the resulting process follows analogously to the derivation of (\ref{eq:svar_sub}) and can be written as
\begin{align}
\x_t &= \F \tilde{\x}_{t-1} + \Ll \tilde{\e}_t \label{eq:mf_rep}, 
\end{align}
where $\tilde{\x}_{t-1}$ are \emph{observed} lags of the series, $\F$ is a function of elements of $A$, and $L$ follows analogously to the subsampled case using elements of $A$ pre-multiplying elements of $C$.  As in the subsampled case, we refer to a parameterization of a mixed\verb+-+frequency structural model as $(\A, \C, p_e;{\bf k})$, where ${\bf k}$ is now a $p$-vector of sampling rates.

The similar form of (\ref{eq:svar_sub}) and (\ref{eq:mf_rep}) suggests similar identifiability results hold.  However, not accounting for subsampling in the mixed\verb+-+frequency setting, Fig.~\ref{sampling_types}(c), not only leads to the kind of mistaken inferences discussed above, but also to further mistakes unique to the mixed\verb+-+frequency case; see \citet{tank2019biometrika} for examples.

While both lagged Granger causality and instantaneous structural interactions are confounded by subsampling and mixed frequency settings, \citet{tank2019biometrika} show that when accounting for this structure, we may, under some conditions, still estimate the $\A$ and $\C$ matrices of the underlying process directly from the subsampled or mixed frequency data. See Theorem~\ref{mixed_corr_theorem}. The identifiability of $\A$ and $\C$ relies on a set of assumptions outlined below.

\begin{assumption} \label{A1}
 $x_t$ is stationary so that all singular values of $\A$ have modulus less than one; 
 \end{assumption}
 \begin{assumption} \label{A2}
the distributions $p_{e_j}$ are distinct for each $j$ after rescaling $e_j$ by any non-zero scale factor, their characteristic functions are all analytic, or they are all non-vanishing, and none of them has an exponent factor with polynomial of degree at least two;
\end{assumption}
\begin{assumption} \label{A3}
all $p_{e_j}$ are asymmetric.
\end{assumption}
\begin{assumption} \label{A4}
the variance of each $p_{e_j}$ is equal to one, i.e., $\Lambda = I_p$;
\end{assumption}
\begin{assumption} \label{A5}
the matrix $\C$ is full rank.
\end{assumption}

\begin{theorem} \label{mixed_corr_theorem} \citep{tank2019biometrika}
 Suppose that $e_{tj}$ are all non-Gaussian and independent, and the data $\tilde{\x}_t$ are generated by (\ref{eq:svar_f}) with representation $(\A, \C,p_e;{\bf k})$. Assume that the process also admits another mixed frequency subsampling representation $(\A', \C',p'_e;{\bf k})$. In the pure subsampling case, $k_j=k$ for all $j$. If Assumptions \ref{A1}, \ref{A2} and \ref{A4} hold, then
 
(a) $\C$ is equal to $\C'$ up to permutation of columns and scaling of columns by $1$ or $-1$, that is $\C' = \C P$ where $P$ is a scaled permutation matrix with $1$ or $-1$ elements. This implies $\Sigma = \C\C^\top = \C' \C'^\top = \Sigma'$;

(b) (\emph{mixed-frequency only}) If $\C$ is lower triangular with positive diagonals, i.e. the instantaneous interactions follow a directed acyclic graph, and if for all $i$ there exists a $j$ such that any multiple of $k_i$ is $1$ smaller than some multiple of $k_j$ with $A_{j:} C_{:i} \neq 0$, then $\A = \A'$. \label{2}

(c) If Assumptions \ref{A3} and \ref{A5} also hold, then $\A = \A'$.
 \end{theorem}

Theorem~\ref{mixed_corr_theorem} demonstrates that identifiability of structural models still holds for mixed\verb+-+frequency series with subsampling under non\verb+-+Gaussian errors. The mixed-frequency setting provides additional information to resolve parameter ambiguities in the non-Gaussian setting. Specifically, $A_{ij}$ is identifiable if there is one time step difference between when series $x_j$ and $x_i$ are sampled. This information can be used to resolve sign ambiguities in columns of $\A$, which leads to statement (b) in Theorem~\ref{mixed_corr_theorem}. This result applies directly to the standard mixed\verb+-+frequency setting \citep{anderson:2015,schorfheide:2015} where one series is observed at every time step, as in Fig.~\ref{sampling_types}(b). It also applies to case (d), since there exist time steps where one series is observed one time step before another series.

In the case of subsampling, if the instantaneous causal effects follow a directed acyclic graph, the structure can be identified without any prior information about causal ordering of the variables. 
 \begin{corollary} \citep{tank2019biometrika} \label{corcor}
 If Assumptions \ref{A1}, \ref{A2}, and \ref{A4} hold and the true structural process corresponds to a directed acyclic graph $G$, that is, it has a lower triangular structural matrix $\C$ with positive diagonals, and it admits another representation with structural matrix $\C'$, then $\C = \C'$. Hence the structure of $G$ is identifiable without prior specification of the causal ordering of $G$. 
 \end{corollary}
 
Together, Theorem~\ref{mixed_corr_theorem} and Corollary~\ref{corcor} imply that when the shocks, $\e_t$, are independent and asymmetric, a complete causal diagram of the lagged and the instantaneous effects is fully identifiable from the subsampled time series, $\tilde{\bf X}$. 

To estimate Granger causality from subsampled and mixed\verb+-+frequency time series, \citet{tank2019biometrika} model the non-Gaussian errors of the SVAR as a mixture of Gaussian distributions with $m$ components. The authors develop an expectation-maximization algorithm for joint estimation of the full set of parameters based only on the observed subsampled and mixed\verb+-+frequency data $\tilde{\bf X}$. The method is the same for all scenarios, Fig.~\ref{sampling_types}(a)-(d).

\section{CONCLUSION}\label{sec:conclusion}

In the first part of this paper, we briefly reviewed classical approaches to Granger causality, mentioned some of their applications and discussed their shortcomings. These shortcomings are primarily due to the restrictive (and unattainable) assumptions that are needed in order to infer causal effects from time series data, which was the original premise of Granger causality. They are also due to the limitations of simple approaches that were historically used to investigate Granger causal relations. 

In the second part of the paper, we discussed recent efforts to relax some of the assumptions made by classical approaches and/or generalize their applicability. This includes investigating Granger causal relations among a large set of variables, automatic lag selection, accounting for non-stationarity, developing flexible methods for non-Gaussian and non-continuous observations and attempts to account for differences between the true causal time scale and the frequency of the observed data. These recent developments have expanded the application domains of Granger causality and offer new opportunities for investigating interactions among components of complex systems with the goal of gaining a systems perspective to their joint behavior. 

In spite of recent progress, there is still much more work to be done in this area. Even when not trying to infer causal effects, we would ideally need flexible nonparametric approaches that handle many observed time series while accounting for unmeasured variables and allowing for non-stationarity. However, despite these limitations, emerging data, especially those obtained from interventions over time and perturbations to the system's state, offer new opportunities for discovering causal effect of variables on each other. At minimum, these new data and continued developments in this area can help researchers take the first step towards causal inference by restricting the set of possible causal hypotheses. We believe this area will continue to be an active area of research.

\bibliography{GrangerRefs,alexreferences}
\end{document}